\documentclass[prd,twocolumn,aps,amsmath,nofootinbib,superscriptaddress]{revtex4}
\usepackage{graphicx}
\usepackage{bm}
\usepackage{epsfig}
\usepackage{amsmath}

  \newcount\hour \newcount\minute
  \hour=\time \divide \hour by 60
  \minute=\time
  \newcommand{\mydate}{\ \today \ - \number\hour :\ifnum \minute<10 0\fi 
\number\minute}

\def\Bslash{B\!\!\!\!\slash}
\def\Dslash{D\!\!\!\!\slash}

\def\nslash{n\!\!\!\slash}
\def\bnslash{\bar n\!\!\!\slash}

\def\OMIT#1{}

\newcommand{\nn}{\nonumber} 

\newcommand{\bn}{{\bar n}}
\newcommand{\bea}{\begin{eqnarray}}
\newcommand{\eea}{\end{eqnarray}}

\newcommand{\bnP}{\bar {\cal P}}

\newcommand{\cP}{{\cal P}}

\newcommand{\mcdot}{\!\cdot\!}

\def\lqcd{\Lambda_{\rm QCD}}

\begin{document}


\preprint{ \vbox{\hbox{MIT-CTP 3688} \hbox{LBNL-58941} \hbox{hep-ph/0510241}  }}

\title{SCET Analysis of $B\to K\pi$, $B\to K\bar K$, and $B\to \pi\pi$
  Decays\vspace{0.3cm}}

\author{Christian W.~Bauer}
\affiliation{Ernest Orlando Lawrence Berkeley National Laboratory, University of California, Berkeley, CA 94720}
\author{Ira Z.~Rothstein}
\affiliation{Department of Physics, Carnegie Mellon University,
        Pittsburgh, PA 15213}
\author{Iain W.~Stewart\vspace{0.4cm}}
\affiliation{Center for Theoretical Physics, Laboratory for Nuclear Science,
  Massachusetts Institute of Technology,\\ 
   Cambridge, MA 02139\vspace{0.3cm}}

\begin{abstract}
  
  $B \to K \pi$ and related decays are studied in the heavy quark limit of QCD
  using the soft collinear effective theory (SCET). We focus on results that
  follow solely from integrating out the scale $m_b$, without expanding the
  amplitudes for the physics at smaller scales such as
  $\alpha_s(\sqrt{E_\pi\Lambda_{\rm QCD}})$.  The reduction in the number of
  hadronic parameters in SCET leads to multiple predictions without the need of
  SU(3).  We find that the CP-asymmetry in $B^-\to \pi^0 K^-$ should have a
  similar magnitude and the same sign as the well measured asymmetry in $\bar
  B^0 \to\pi^+ K^-$. Our prediction for ${\rm Br}(K^+\pi^-)$ exceeds the current
  experimental value at the $2 \sigma$ level.  We also use our results to
  determine the corrections to the Lipkin and CP-asymmetry sum rules in the
  standard model and find them to be quite small, thus sharpening their utility
  as a tool to look for new physics.

\end{abstract}

\maketitle

\section{Introduction} 
\label{sec:introduction}
Two body nonleptonic decays are the most widely used processes to study CP
violation in the $B$ system. Due to the large mass of the $B$-meson there is a
plethora of open channels, each of which provides unique ways for testing the
consistency of the standard model.  For each channel observables include the CP
averaged branching ratios ({\rm Br}), direct CP asymmetry ($A_{CP}=-C$), and for
certain neutral $B$ decays, the time dependent CP asymmetry ($S$).  For the
decays we are interested in
\begin{align} \label{obs}
 & {\rm Br} \equiv \frac{1}{\Gamma_B} \frac{s|\vec p|}{8\pi m_B^2} 
   \Big( \frac{|{A}|^2 + |\overline A|^2}{2} \Big) \,, 
  \quad \lambda_{ CP}= \frac{q}{p}\, \frac{\overline A}{A},\\[5pt]
 & A_{\rm CP} \equiv 
    \frac{|{A}|^2 - |\overline A|^2}{|{A}|^2 + |\overline A|^2} \,,
  \quad S = \frac{2 \, {\rm Im}(\lambda_{CP})}{1+|{\lambda_{ CP}}|^2}\,,
   \nn\\[5pt]
 & \frac{\Gamma_{B^0}(t) - \Gamma_{\bar B^0}(t)}
   {\Gamma_{B^0}(t) + \Gamma_{\bar B^0}(t)}
    \equiv -S \sin(\Delta m\, t) + C \cos(\Delta m\, t)
  \nn \,,
\end{align}
where $A$ is the amplitude of the decay process $A=A(\bar B\to M_1 M_2)$,
$\overline A$ is the amplitude for CP-conjugate process, and $q/p$ is the mixing
parameter for $B^0-\bar B^0$ and/or $K^0-\bar K^0$ mixing.  The other parameters
in Eq.(\ref{obs}) are $|\vec p|$, the final meson momentum in the $B$ rest
frame, $s$, a possible identical particle symmetry factor, and $\Delta m$, the
difference between mass eigenstates in the neutral $B$ two-state system.

Using the unitarity of the CKM matrix to remove top-quark CKM elements, the
amplitude for any decay can be written with the CKM elements factored out as
\begin{eqnarray} \label{Auc}
A = \lambda_u^{(f)} A_u + \lambda_c^{(f)} A_c\,,
\end{eqnarray}
where $\lambda_p^{(f)} = V_{pb}^* V_{pf}$.  Theoretical predictions for the
observables in (\ref{obs}) are often hampered by our ability to calculate
$A_{u,c}$. In general the CP-asymmetries depend on the ratio of amplitudes
$|A_u/A_c|$ and their relative strong phase $\delta$. In fact $A_{\rm CP}\propto
\sin(\delta)$, and so non-negligible strong dynamics are required for the
existence of a direct CP asymmetry.

The parameters $A_u$ and $A_c$ are in principle different for each decay
channel. In order to accurately determine $A_u$ and $A_c$ we need model
independent methods to handle the strong dynamics in these decays.  All such
methods involve systematic expansions of QCD in ratios of quark masses and the
scale $\Lambda\simeq \Lambda_{\rm QCD}$ associated with hadronization. This
includes flavor symmetries for the light quarks, SU(2) and SU(3), from
$m_q/\Lambda\ll 1$, as well as expansions for the heavy $b$-quark from
$\Lambda/m_b \ll 1$. For nonleptonic decays to two light mesons with energies
$E_m\sim m_b/2$, kinematics implies that we must also expand in $\Lambda/E_M \ll
1$.  A formalism for systematically expanding QCD in this fashion is the
soft-collinear effective theory (SCET)~\cite{SCET}.  In nonleptonic B-decays the
expected accuracy of these expansions are
\begin{align} \label{expn}
  & {\rm SU(2)} & \frac{m_{u,d}}{\Lambda} \sim 0.03  \ll 1 \,,\\
  & {\rm SU(3)} & \frac{m_{u,d}}{\Lambda} \ll 1 ,\ 
  \frac{m_{s}}{\Lambda} \sim 0.3  \ll 1 \,, \nn \\
  & {\rm SCET} & \frac{\Lambda}{E_M} \sim \frac{2\Lambda}{m_b}
    \sim 0.2 \ll 1
  \,. \nn
\end{align}
The flavor symmetries SU(2) and SU(3) provide amplitude relations between
different nonleptonic channels, thereby reducing the number of hadronic
parameters. The expansion in $\Lambda/m_b\sim \Lambda/E_M$ also reduces the
number of hadronic parameters. In this case the expansion yields factorization
theorems for the amplitudes in terms of moments of universal hadronic functions.

In this paper we study standard model predictions for $B\to K\pi$, $K\bar K$,
and $\pi\pi$ decays. These channels provide 25 observables, of which 19 have
been measured or bounded as summarized in Table~\ref{dataall}.  We make use of
the expansions in Eq.~(\ref{expn}), focusing on SCET. Our goal is to quantify the
extent to which the current data agrees or disagrees with the standard model in
the presence of hadronic uncertainties, and to provide a roadmap for looking for
deviations in future precision measurements of these decays.
\begin{table}[t!]
\begin{tabular}{l|ccc}
 & ${\rm Br}\times 10^6$ & $A_{\rm CP}=-C$ & $S$ \\\hline
$\pi^+ \pi^- $ &\ $5.0\pm 0.4$ \  & \ $0.37\pm  0.10$ \  &\  $-0.50\pm 0.12$ \ \\
$\pi^0 \pi^0$ & $1.45\pm 0.29$ & $0.28\pm  0.40$ & \\
$\pi^+ \pi^0$ & $5.5\pm 0.6$ & $0.01 \pm 0.06$ & $-$ \\\hline
$\pi^- \bar K^0 $ &\ $24.1\pm 1.3$ \  & \ $-0.02\pm  0.04$ \   
 & $-$\ \\
$\pi^0 K^-$ & $12.1\pm 0.8$ & $0.04\pm  0.04$ & $-$\\
$\pi^+ K^-$ & $18.9\pm 0.7$ & $-0.115\pm  0.018$ & $-$ \\
$\pi^0 \bar K^0$ & $11.5\pm 1.0$ & $-0.02 \pm 0.13$ & $0.31 \pm 0.26$ \\\hline
$K^+K^-$ & $0.06\pm 0.12 $&\ & \   \ \\
$K^0 \bar K^0$&$0.96 \pm 0.25$&\ & \ \\
$\bar K^0 K^-$&$1.2\pm0.3$&\ & $-$ \\
\hline
\end{tabular}
{\caption {Current  $B \to \pi\pi$, $K\pi$, and $K\bar K$ 
data~\cite{HFAG,Babar,Belle,Cleo2,CDF}. The $S$ for $\pi K$ is $S(\pi^0 K_S)$.
}\label{dataall}}
\end{table}

The SU(2) isospin symmetry is known to hold to a few percent accuracy, and thus
almost every analysis of nonleptonic decays exploits isospin symmetry.
(Electroweak penguin contributions are simply $\Delta I=1/2$ and $\Delta I=3/2$
weak operators, and are not what we mean by isospin violation.)  Methods for
determining or bounding $\alpha$ (or $\gamma$) using isospin have been discussed
in~\cite{GL,isospinbound} and are actively used in $B\to\pi\pi$ and $B\to
\rho\rho$ decays. In $B\to\rho\rho$ this yields $\alpha_{\rho\rho}= 96^\circ \pm
13^\circ$~\cite{HFAG}. For $B\to\pi\pi$ this analysis has significantly larger
errors, since the $A_c$ amplitudes are larger and the asymmetry $C(\pi^0\pi^0)$
is not yet measured well enough to constrain the hadronic parameters. Isospin
violating effects have been studied in~\cite{isospinbreaking}. For $B\to K\pi$
and $B\to K\bar K$ an SU(2) analysis is not fruitful since there are more
isospin parameters than there are measurements, so further information about the
hadronic parameters is mandatory.

In $B\to \pi\pi$, even if $C(\pi^0\pi^0)$ were known precisely it would still be
important to have more information about the amplitudes $A_u$ and $A_c$ than
isospin provides.  For example, isospin allows us to test whether
$\gamma_{\pi\pi}$ differs from the value obtained by global
fits~\cite{CKMfitter,UTfit},
\begin{align}
  \gamma_{\rm\, global}^{\rm\, CKMfitter} 
     &= 58.6^\circ {}^{+6.8^\circ}_{-5.9^\circ} 
  \,,\nn\\
  \gamma_{\rm\, global}^{\rm\, UTfit} 
     &= 57.9^\circ \pm 7.4^\circ \,.
\end{align}
However, a deviation in $\gamma$ is not the only way that new physics can appear
in $B\to\pi\pi$ decays. Simply fitting the full set of SU(2) amplitudes can
parameterize away a source of new physics. For example, Ref.~\cite{Baek:2005cg}
has argued that it is impossible to see new physics in the $(\pi\pi)_{I=0}$
amplitudes in an isospin based fit. Thus, it is important to consider the
additional information provided by SU(3) or factorization, since this allows us
to make additional tests of the standard model.  The expansion parameters here
are larger, and so for these analyses it becomes much more important to properly
assess the theoretical uncertainties in order to interpret the data.

The analysis of  $B\to K\pi$ decays has a   rich history in the  standard model,
provoked by the CLEO  measurements~\cite{CLEO} that indicated that these  decays
are   dominated by  penguin  amplitudes  that  were  larger  than  expected. The
dominance  by loop effects makes  these decays an   ideal place to  look for new
physics effects.   Some    recent new    physics  analyses   can   be found   in
Refs.~\cite{Kpinewphysics}.  This literature is divided on  whether or not there
are hints for new physics in these decays. The main obstacle is the assessment
of the uncertainty of the standard model predictions from hadronic interactions.

Several standard model analyses based on the limit $m_s/\Lambda \ll 1$ (ie SU(3)
symmetry) have been reported
recently~\cite{SU3gronau,SU3buras,SU3Li,Suprun,SU3Wu,SU3other} (see
also~\cite{su3,graphical2,burasfleischer} for earlier work).  In the $\Delta S = 1$ decays the electroweak
penguin amplitudes can not be neglected, since they are enhanced by CKM factors.
Unfortunately the number of precise measurements makes it necessary to introduce
additional ``dynamical assumptions'' to reduce the number of hadronic parameters
beyond those in SU(3).  In some cases efforts are made to estimate a subset of
the SU(3) violating effects to further reduce the uncertainty.  The dynamical
assumptions rely on additional knowledge of the strong matrix elements and in
the past were motivated by naive factorization or the large $N_c$ limit of QCD.
Our current understanding of the true nature of factorization in QCD allows some
of these assumptions to be justified by the $\Lambda/E_M$ expansion. However, it
should be noted that a priori there is no reason to prefer these factorization
predictions to others that follow from the $\Lambda/m_b$ expansion (such as the
prediction that certain strong phases are small).
 
In Ref.~\cite{SU3gronau} a $\chi^2$-fit was performed with $\gamma$ as a fit
parameter, including decays to $\eta$ and $\eta'$. The result $\gamma=61^\circ
\pm 11^\circ$ agrees well with global CKM fits.  Here evidence for deviations
from the standard model would show up as large contributions to the $\chi^2$.
The most recent analysis~\cite{Suprun} has ${\rm Br}(K^+\pi^-)$, ${\rm
  Br}(K^0\pi^0)$, and $A_{\rm CP}(K^0\pi^0)$ contributing $\Delta \chi^2 = 2.7$,
$5.9$, and $2.9$ respectively, giving some hints for possible deviations from
the standard model.  Ref.~\cite{SU3buras} extracted hadronic paramters from
$B\to\pi\pi$ decays, and used these results together with SU(3) and the neglect
of exchange, penguin annihilation, and all electroweak penguin topologies except
for the tree to make predictions for $B\to K\pi$ and $B\to K\bar K$ decays. They
find large annihilation amplitudes, a large phase and magnitude for an amplitude
ratio $\tilde C/\tilde T$ which is interpretted as large $P_{ut}$ penguin
amplitudes. The deviation of ${\rm Br}(K^+\pi^-)/{\rm Br}(\bar K^0\pi^0)$ from
standard model expectations was interpreted as evidence for new physics in
electroweak penguins.

There has been tremendous progress over the last few years in understanding
charmless two-body, non-leptonic $B$ decays in the heavy quark limit of
QCD~\cite{QCDF,PQCD,charmingpenguins,earlier,pipiChay,bprs,Bauer:2001cu,Mantry:2003uz,diff1,diff2,FH,pQCDKpi,BW,BS,Lee,Kagan}.
In this limit one can prove factorization theorems of the matrix elements
describing the strong dynamics in the decay into simpler structures such as
light cone distribution amplitudes of the mesons and matrix elements describing
a heavy to light transition~\cite{QCDF} (for earlier work see
Refs.~\cite{earlier}). It is very important that these results are obtained from
a systematic expansion in powers of $\lqcd/m_b$. The development of
soft-collinear effective theory (SCET)~\cite{SCET} allowed these decays to be
treated in the framework of effective theories, clarifying the separation of
scales in the problem, and allowing factorization to be generalized to all
orders in $\alpha_s$. In Ref.~\cite{Bauer:2001cu} a proof of factorization was
given for $B\to DM^-$ type decays. Power corrections can also be investigated
with SCET and in Ref.~\cite{Mantry:2003uz} a factorization theorem was proven
for the color-suppressed $\bar B^0\to D^0 M^0$ decays, and extended to
isosinglet light mesons in Ref.~\cite{Blechman:2004vc}. Predictions from these
results agree quite well with the available data, in particular the prediction
of equal rates and strong phase shift for $D$ and $D^*$ channels.

Factorization for $B\to M_1 M_2$ decays involves three distinct distance scales
$m_b^2 \gg E_M\Lambda \gg \Lambda^2$. For $B\to M_1 M_2$ decays a factorization
theorem was proposed by Beneke, Buchalla, Neubert and Sachrajda~\cite{QCDF},
often referred to as the QCDF result in the literature. Another proposal is a
factorization formula which depends on transverse momenta, which is referred to
as PQCD~\cite{PQCD}. The factorization theorem derived using
SCET~\cite{bprs,pipiChay} agrees with the structure of the QCDF proposal if
perturbation theory is applied at the scales $m_b^2$ and $m_b\Lambda$.  (QCDF
treats the $c\bar c$ penguins perturbatively, while in our analysis they are
left as a perturbative contribution plus an unfactorized large ${\cal O}(v)$
term.)  Due to the charm mass scale the identification of a convergent
expansion for the $c\bar c$ penguins remains
unclear~\cite{charmingpenguins,FH,diff1,diff2}. For further discussion
see~\cite{diff1,diff2}.)  The SCET result improved the factorization formula by
generalizing it to allow each of the scales $m_b^2$, $E_M\Lambda$, and
$\Lambda^2_{\rm QCD}$ to be discussed independently.  In particular, it was
possible to show that a reduced set of universal parameters for these decays can
already be defined after integrating out the scale $m_b^2$~\cite{bprs}, opening
up the ability to make predictions for nonleptonic decays without requiring an
expansion in $\alpha_s(\sqrt{E\Lambda})$. (If the $m_b^2$ and $m_b\Lambda$
scales were separated in pQCD then this same result would be found for this
first stage of factorization.) As a secondary step, additional predictions can
be explored by doing a further expansion in $\alpha_s$ at the intermediate
scale.  The expense of the second expansion comes in principle with the benefit
of a further reduction in the number of hadronic parameters and additional
universality. In this paper we will explore the implications the first step of
factorization has for $B\to K\pi$ decays.

There are several ways results from factorization can be used to analyze the
data depending on i) whether perturbation theory is used at the intermediate
$E_M\Lambda$ scale as mentioned above, and ii) whether light-cone sum rules,
models, or data is used to determine the hadronic parameters.  In the
QCDF~\cite{QCDF} and PQCD~\cite{PQCD} analyses perturbation theory is used at
the scale $E_M\Lambda$ and light-cone sum rules~\cite{Ball,Koj} or simple
estimates were used for numerical values of most of the hadronic parameters.
Nonleptonic decay have also been studied with light-cone sum rules~\cite{Khodj}.
With this input, all nonleptonic observables can be predicted and confronted
with the experimental data. In both QCDF and PQCD a subset of power corrections
are identified, parameterized in terms of new unknowns, and included in the
numerical analysis. These power corrections are crucial to get reasonable agreement
with the data. In these analyses
it is sometimes difficult to distinguish between the model independent
predictions from the heavy quark limit and the model dependent input from
hadronic parameters. Ciucchini et al. have argued that so called charming
penguins could be larger than expected and include unknowns to parameterize
these effects~\cite{charmingpenguins}.  Fitting the hadronic parameters to
non-leptonic data in some channels and using the results to make predictions for
other channels, as we advocate in this paper, has the advantage of avoiding model
dependent input. Fits in QCDF have been performed in~\cite{CKMfitter}. So far 
restrictions on the size of leading and subleading hadronic parameters necessary 
to guarantee convergence have not been explored. Other fits based purely on 
isospin symmetry have been explored in~\cite{isofits,bprs}.

In Ref.~\cite{bprs,gammafit,GHLP} the factorization theorem was used in a
different way, focusing on $B\to \pi\pi$ decays. Here perturbation theory was
only used at the $m_b^2$ scale and fits to nonleptonic data were performed for
the hadronic parameters in the LO factorization theorem. The problematic
contributions from charm-quark penguins were treated using only isospin
symmetry.  (This is also a good approach if power corrections spoil the
expansion for this observable. Note that it avoids expanding the amplitude which
has possible contamination from ``chirally enhanced'' power
corrections~\cite{QCDF}.)  Here we continue this program for $B\to K\pi$ and
$B\to K\bar K$ decays (along with there comparison with $B\to \pi\pi$).  For
simplicity we refer to this as an ``SCET'' analysis, although it should be
emphasized that other approaches to using the SCET-factorization theorem are
possible.  A key utility of factorization for nonleptonic decays is that the
$\Lambda/E$ and $\alpha_s(m_b)$ expansions are systematic and give us a method
to estimate the theory uncertainty. Based on these uncertainties we investigate
if the theory at leading order is able to explain the observed data.  When
deviations are found there are several possible explanations, all of which are
interesting: either the expansions inherent in the theoretical analysis are
suspect, or there are statistical fluctuations in the data, or we are seeing
first hints of physics beyond the standard model.

This paper is organized as follows: In section~\ref{sec:theory} we discuss the
theory input required to describe the decays of a $B$ meson to two light
pseudoscalar mesons. We briefly review the electroweak Hamiltonian at $\mu=m_b$
and then we discuss the counting of the number of parameters required to
describe these decays using SU(2), SU(3), and SCET analyses.  We finish this
section by giving a general parameterization of the decay amplitudes in SU(2).
(In the appendix we give the relations between our parameters and the graphical
amplitudes~\cite{graphical1,graphical2,graphical3}.) In
section~\ref{sec:SCETtheory} we give the expressions of the decay amplitudes in
SCET. We begin by giving the general expressions at leading order in the power
expansion, but correct to all orders in $\alpha_s$ and comment about new
information that arises from combining these SCET relations with the SU(3)
flavor symmetry. We then use the Wilson coefficients at leading order in
$\alpha_s(m_b)$ and give expressions for the decay amplitudes at that order. We
finish this section with a discussion of our estimate of the uncertainties which
arise from unknown ${\cal O}(\alpha_s(m_b))$ and ${\cal O}(\lqcd/E)$
corrections. A detailed discussion of the implications of the SCET results is
given in section~\ref{sec:SCETimplications}. We emphasize that within factorization 
the ratios of
color suppressed and color allowed amplitudes ($C/T$ and $EW^C/EW^T$) can
naturally be of order unity at LO in the power counting, contrary to
conventional wisdom~\cite{bprs}.  We also perform an error analysis for the
Lipkin and CP-sum rules in $B\to K\pi$ decays, and discuss predictions for the
relative signs of the $CP$ asymmetries. We then review the information one can
obtain from only the decays $B \to \pi \pi$, before we discuss in detail the
implications of the SCET analysis for the decays $B \to K \pi$ and $B \to KK$.

\section{Theory Input}
\label{sec:theory}
\subsection{The electroweak Hamiltonian}
\label{sec:hamiltonian}
The electroweak Hamiltonian describing $\Delta b=1$ transitions $b\to f$ is
given by
\begin{eqnarray} \label{Hw}
 H_W = \frac{G_F}{\sqrt{2}} \sum_{p=u,c} \lambda_p^{(f)}
 \Big( C_1 O_1^p + C_2 O_2^p 
  +\!\!\! \sum_{i=3}^{10,7\gamma,8g}\!\! C_i O_i \Big),
\end{eqnarray}
where the CKM factor is $\lambda_p^{(f)} = V_{pb} V^*_{pf}$.  The standard basis
of operators are (with $O^p_{1}\leftrightarrow O^p_{2}$ relative
to~\cite{fullWilson}) 
\begin{eqnarray}\label{fullops}
 O_1^p \!\! &=&\!\! (\overline{p} b)_{V\!-\!A}
  (\overline{f} p)_{V\!-\!A}, \ \
 O_2^p = (\overline{p}_{\beta} b_{\alpha})_{V\!-\!A}
  (\overline{f}_{\alpha} p_{\beta})_{V\!-\!A}, \nonumber \\
 O_{3,4} \!\! &=& \!\! \big\{ (\overline{p} b)_{V\!-\!A}
  (\overline{q} q)_{V\! - \!A}\,, (\overline{f}_{\beta} b_{\alpha})_{V\!-\!A}
  (\overline{q}_{\alpha} q_{\beta})_{V\! - \!A} \big\}, \nonumber \\
 O_{5,6} \!\! &=& \!\! \big\{ (\overline{f} b)_{V\!-\!A}
  (\overline{q} q)_{V\! + \!A}\,, (\overline{f}_{\beta} b_{\alpha})_{V\!-\!A}
  (\overline{q}_{\alpha} q_{\beta})_{V\! + \!A} \big\}, \nonumber \\
 O_{7,8} \!\! &=& \frac{3e_q}{2}\!\! \big\{ (\overline{f} b)_{V\!-\!A}
  (\overline{q} q)_{V\! + \!A}\,, (\overline{f}_{\beta} b_{\alpha})_{V\!-\!A}
  (\overline{q}_{\alpha} q_{\beta})_{V\! + \!A} \big\}, \nonumber \\
 O_{9,10} \!\! &=& \frac{3e_q}{2}\!\! \big\{ (\overline{f} b)_{V\!-\!A}
  (\overline{q} q)_{V\! - \!A}\,, (\overline{f}_{\beta} b_{\alpha})_{V\!-\!A}
  (\overline{q}_{\alpha} q_{\beta})_{V\! - \!A} \big\}, \nonumber \\
 O_{7\gamma,8g} \!\! &=&\!\!  -\frac{m_b}{8\pi^2}\ \overline{f}\, \sigma^{\mu\nu}
  \{e F_{\mu\nu},g G_{\mu\nu}^a T^a\} (1\!+\! \gamma_5)  b \,.
\end{eqnarray}
Here the sum over $q=u,d,s,c,b$ is implicit, $\alpha, \beta$ are color indices
and $e_q$ are electric charges. The $\Delta S=0$ and $\Delta S=1$ effective
Hamiltonian is obtained by setting $f=d$ and $f=s$ in
Eqs.~(\ref{Hw},\ref{fullops}), respectively.  The Wilson coefficients are known
to NLL order~\cite{fullWilson}. At LL order taking $\alpha_s(m_Z)=0.118$,
$m_t=174.3$, and $m_b=4.8\,{\rm GeV}$ gives $C_{7\gamma}(m_b)=-0.316$,
$C_{8g}(m_b) =-0.149$ and
\begin{eqnarray}
 && 
C_{1-10}(m_b) = \{
  1.107\,, 
  -.249\,,
  .011\,,
 -.026\,, 
  .008\,, 
 -.031 \,, 
  \nn\\
 && \ \ 
  4.9 \!\times\! 10^{-4} \,,
  4.6 \!\times\! 10^{-4} \,,
  -9.8 \!\times\! 10^{-3} \,,
  1.9 \!\times\! 10^{-3} \} \,.
\end{eqnarray}

Below the scale $\mu \sim m_b$ one can integrate out the $b \bar b$ pairs in the
operators $O_{3-10}$. The remaining operators have only one $b$-quark field, and
sums over light quarks $q=u,d,s,c$.  This gives rise to a threshold
correction to the Wilson coefficients,
\begin{eqnarray}
 C_i^-(m_b) = C_i^+(m_b) \Big[1 + \frac{\alpha_s(m_b)}{4\pi} \delta r_s^T +
 \frac{\alpha}{4 \pi} \delta r_c^T \Big]\,,
\end{eqnarray}
where $C^+$ and $C^-$ are the Wilson coefficients with and without dynamical $b$
quarks, and $\delta r_s^T$ and $\delta r_c^T$ are given in Eqs.~(VII.31)
and~(VII.32) of~\cite{fullWilson}. This changes the numerical values of the
Wilson coefficients by less than 2\%. Integrating out dynamical b quarks allows
for additional simplifications for the electroweak penguin operators, since now
for the flavor structure we have
\begin{align}
 \frac{3}{2}\, e_q (\bar f b)(\bar q q) 
 &= \frac{1}{2}\, (\bar f b) ( 2 u \bar u-  d \bar d -  s \bar s + 2c \bar c)  
  \\
 &\!\!\!\!
= \frac{3}{2} \, (\bar f b) ( u \bar u) + \frac{3}{2}(\bar f b) ( c \bar c) 
    - \frac12 \!\!\!\! \sum_{q=u,d,s,c}\!\!\! (\bar f b) (q\bar q) 
 \,.\nn
\end{align}

The operators $O_9$ and $O_{10}$ have the regular $(V-A)\times(V-A)$ Dirac
structure, and can therefore be written as linear combinations of the operators
$O_{1-4}$,
\begin{align} \label{O9O10}
   O_9 &= \frac32\, O_2^u +  \frac32\, O_2^c - \frac12\, O_3 \,, \\
   O_{10} &= \frac32\, O_1^u +  \frac32\, O_1^c - \frac12\, O_4 \,. \nn
\end{align}
This is not possible for the operators $O_7$ and $O_8$, which have
$(V-A)\times(V+A)$ Dirac structure. Thus, integrating out the dynamical $b$
quarks removes two operators from the basis. To completely integrate out the
dynamics at the scale $m_b$ we must match onto operators in SCET, as discussed
in section~\ref{sec:SCETtheory} below.

\subsection{Counting of Parameters}
\label{sec:counting}

Without any theoretical input, there are 4 real hadronic parameters for each
decay mode (one complex amplitude for each CKM structure) minus one overall
strong phase.  In addition, there are the weak CP violating phases that we want
to determine. For $B \to \pi \pi$ decays there are a total of 11 hadronic
parameters, while in $B \to K \pi$ decays there are 15 hadronic parameters.

Using isospin, the number of parameters is reduced. Isospin gives one amplitude
relation for both the $\pi \pi$ and the $K \pi$ system, thus eliminating 4
hadronic parameters in each system (two complex amplitudes for each CKM
structure). This leaves 7 hadronic parameters for $B \to \pi \pi$ and 11 for $B
\to K \pi$. An alternative way to count the number of parameters is to construct
the reduced matrix elements in SU(2). The electroweak Hamiltonian mediating the
decays $B \to \pi \pi$ has up to three light up or down quarks. Thus, the
operator is either $\Delta I = 1/2$ or $\Delta I = 3/2$. The two pions are
either in an $I=0$ or $I=2$ state (the $I=1$ state is ruled out by Bose
symmetry). This leaves 2 reduced matrix elements for each CKM structure, $\langle {\bf 0} || {\bf 1/2} || {\bf 1/2} \rangle$ 
and $\langle {\bf
  2} || {\bf 3/2} || {\bf 1/2} \rangle$.  For $B \to K \pi$ decays the
electroweak Hamiltonian has either $\Delta I = 0$ or $\Delta I = 1$. The $K \pi$
system is either in an $I=1/2$ or $I=3/2$ state thus there are three reduced
matrix elements per CKM structure, $\langle {\bf 3/2} || {\bf 1} ||
{\bf 1/2} \rangle$, $\langle {\bf 1/2} || {\bf 1} || {\bf 1/2}
\rangle$ and $\langle {\bf 1/2} || {\bf 0} || {\bf 1/2} \rangle$.
Finally, $K\bar K$ is either an $I=0$ or $I=1$, and there are again three
reduced matrix elements per CKM structure, $\langle {\bf 0} || {\bf 1/2}
|| {\bf 1/2} \rangle$, $ \langle {\bf 1} || {\bf 1/2} || {\bf 1/2}
\rangle$, and $\langle {\bf 1} || {\bf 3/2} || {\bf 1/2} \rangle$.

The SU(3) flavor symmetry relates not only the decays $B \to \pi \pi$ and $B \to
K \pi$, $B  \to K K$, but  also $B \to \pi \eta_8$,  $B \to \eta_8 K$  and $B_s$
decays to  two  light mesons. The  decomposition  of the amplitudes  in terms of
SU(3)     reduced          matrix        elements     can       be      obtained
from~\cite{Zeppenfeld,SavageWise,GrinsteinLebed}. The Hamiltonian can  transform
either as a  ${\bf \overline 3}^s$, ${\bf   \overline 3}^a$, ${\bf  6}$ or ${\bf
  \overline  {15}}$.     Thus, there are   7  reduced  matrix elements   per CKM
structure, $\langle {\bf 1} || {\bf \overline 3}^s || {\bf 3} \rangle$, $\langle
{\bf 1}  || {\bf \overline 3}^a ||  {\bf 3} \rangle$, $\langle   {\bf 8} || {\bf
  \overline 3}^s || {\bf 3} \rangle$, $\langle {\bf 8} || {\bf \overline 3}^a ||
{\bf  3} \rangle$, $\langle  {\bf 8} || {\bf  6}^s || {\bf 3} \rangle$, $\langle
{\bf 8} || {\bf \overline {15} }^s || {\bf 3}  \rangle$ and $\langle {\bf 27} ||
{\bf \overline{15}}^s || {\bf 3} \rangle$.  The  ${\bf \overline 3}^a$ and ${\bf
  \overline 3}^s$ come in a single linear combination so this leaves 20 hadronic
parameters to describe all  these decays minus 1  overall phase (plus additional
parameters for singlets and mixing to properly describe $\eta$ and $\eta'$).  Of
these hadronic parameters, only 15 are required to describe $B  \to \pi \pi$ and
$B \to K \pi$ decays (16 minus an overall phase).  If we add $B \to K K $ decays
then 4  more   paramaters  are needed  (which  are  solely  due to   electroweak
penguins).  This is discussed further in section~\ref{sec:SU3relations}.

\begin{table}
\begin{tabular}{l||c|c|c|c|c|}
& no & &  & SCET & SCET   \\[-2pt]
 & expn. &  \raisebox{1.6ex}[0pt]{SU(2)} 
 & \raisebox{1.6ex}[0pt]{SU(3)} 
 & +SU(2)&+SU(3) \\\hline
$B \to \pi \pi$  & 11 & 7/5 & & 4 &\\ \cline{1-3}\cline{5-5}
$B \to K \pi$  & 15& 11 & \raisebox{1.6ex}[0pt]{15/13} &
    +5(6) & \raisebox{1.6ex}[0pt]{4} \\ \hline
$B \to K \bar K$  & 11 & 11 & +4/0 & +3(4) & +0
\end{tabular}
\caption{
Number of real hadronic parameters from different expansions in QCD. The first column
shows the number of theory inputs with no approximations, while the next columns
show the number of parameters using only SU(2), 
using only SU(3), using SU(2) and SCET, and using SU(3) with SCET.  For the 
cases with two numbers, $\#/\#$, the second follows from the first after 
neglecting the small penguin coefficients, ie setting $C_{7,8}=0$.  In SU(2) 
+ SCET $B\to K\pi$ has 6 parameters, but 1 appears already in $B\to \pi\pi$, 
hence the $+5(6)$. The notation is analogous for the $+3(4)$ for 
$B\to K\bar K$. 
\label{table_parameters}
}
\end{table}
The number of parameters that occur at leading order in different expansions of
QCD are summarized in Table~\ref{table_parameters}, including the SCET
expansion. Here by SCET we mean after factorization at $m_b$ but without using
any information about the factorization at $\sqrt{E\Lambda}$.  The SCET results
are discussed further in section~\ref{sec:SCETtheory}, but we summarize them
here.  The parameters with isospin+SCET are
\begin{align} \label{params}
  & \pi\pi: &\{& \zeta^{B\pi}\!+\! \zeta_J^{B\pi},\beta_\pi  \zeta_J^{B\pi},
    P_{\pi\pi} \} \,, \\
  & K\pi:   &\{& \zeta^{B\pi}\!+\! \zeta_J^{B\pi},\beta_{\bar K}  \zeta_J^{B\pi},
    \zeta^{B\bar K}+\zeta_J^{B\bar K}, \beta_\pi \zeta_J^{B\bar K}, 
     P_{K\pi} \} \,, \nn \\
  & K\bar K: &\{& \zeta^{B\bar K}+ \zeta_J^{B\bar K},\beta_K \zeta_J^{B\bar K}, 
     P_{K\bar K} \} \,. \nn
\end{align}
Here $P_{M_1 M_2}$ are complex penguin amplitudes and the remaining parameters
are real.\footnote{The penguin amplitudes are kept to all orders in
  $\Lambda/m_b$ since so far there is no proof that the charm mass $m_c$ does
  not spoil factorization, with large $\alpha_s(2m_c) v$ contributions competing
  with $\alpha_s(m_b)$ hard-charm loop corrections~\cite{bprs}. This is
  controversial~\cite{diff1,diff2}. Our analysis treats these contributions in
  the most conservative possible manner. } In $B\to \pi\pi$ the moment parameter
$\beta_\pi$ is not linearly independent from the parameters $\zeta^{B\pi}$ and
$\zeta^{B\pi}_J$, and only the product $\beta_\pi \zeta_J^{B\pi}$ was counted as
a parameter.  In any case it is fairly well known from fits to
$\gamma^*\gamma\to\pi^0$~\cite{pigammafit} $3\beta_\pi\equiv \langle
x^{-1}\rangle_\pi \simeq 3.2\pm 0.2$.  In isospin + SCET $B\to K\pi$ has 6
parameters, but the first one listed in (\ref{params}) appears already in $B\to
\pi\pi$, hence the $+5$ in Table~\ref{table_parameters}.  If the ratio
$\beta_K/\beta_\pi$ was known from elsewhere then one more parameter can be
removed for $K\pi$ (leaving +4).  For $B\to K\bar K$ we have $4$ SCET
parameters. One of these appears already in $B\to K\pi$, hence the +3, and if
$\beta_K/\beta_{\bar K}$ is known from other processes it would become $+2$.

Taking SCET + SU(3) we have the additional relations $\zeta^{B\pi} = \zeta^{B K}
= \zeta^{B\bar K}$, $\zeta_J^{B\pi} = \zeta_J^{B K} = \zeta_J^{B\bar K}$,
$\beta_\pi=\beta_K=\beta_{\bar K}$, and $A_{cc}^{\pi\pi}= A_{cc}^{K\pi} =
A_{cc}^{K\bar K}$ which reduces the number of parameters considerably.

Note that there are good indications that the parameters $\zeta^{BM}$ and
$\zeta^{BM}_J$ are positive numbers in the SCET factorization theorem.
($\beta_K$, $\beta_\pi$, $\beta_{\bar K}$ are also positive.) This follows from:
i) the fact that $\zeta^{BM}+\zeta^{BM}_J$ are related to form factors for
heavy-to-light transitions which with a suitable phase convention one expects
are positive for all $q^2$, ii) that $\zeta^{BM}_J$ is positive (from the
relatively safe assumption that radiative corrections at the scale
$\sqrt{E\Lambda}$ do not change the sign of $\zeta_J^{M_1M_2}$ and that
$\zeta_J\propto \beta_\pi\lambda_B>0$), and finally iii) that the fit to
$B\to\pi\pi$ data gives $\zeta^{B\pi},\zeta_J^{B\pi}>0$ so that SU(3) implies
$\zeta^{BK},\zeta_J^{BK}>0$.  We will see that this allows some interesting
predictions to be made even without knowing the exact values of the parameters.

In using the expansions in (\ref{expn}) it is important to keep in mind the
hierarchy of CKM elements, and the rough hierarchy of the Wilson coefficients
\begin{align}
 C_1 \gtrsim C_2 \gg C_{3-6} \gg C_{9,10} \gtrsim C_{7,8} \,.
\end{align}
Some authors attempt to exploit the numerical values of the Wilson coefficients
in the electroweak Hamiltonian to further reduce the number of parameters.  A
common example is the neglect of the coefficients $C_{7,8}$ relative to
$C_{9,10}$. In Eq.~(\ref{O9O10}) the electroweak penguin operators $O_9$ and
$O_{10}$ were written as linear combinations of $O_{1-4}$.  This implies that if
one neglects the electroweak penguin operators $Q_7$ and $Q_8$, then no new
operators are required to describe the EW penguin effects. In some cases this
leads to additional simplifications. One can show that for $B \to \pi \pi$
decays the $\Delta I=3/2$ amplitudes multiplying the CKM structures $\lambda_u$
and $\lambda_c$ are identical~\cite{graphical2,burasfleischer}.  Thus, SU(2) gives one
additional relation between complex amplitudes in the $\pi\pi$ system, reducing
the hadronic parameters to 5. For $B \to K \pi$ decays the operators giving rise
to the $A_{3/2}$ reduced matrix elements are identical for the $\lambda_u$ and
$\lambda_c$ CKM structures only if SU(3) flavor symmetry is
used~\cite{NeubertRosner}. Thus, for these decays two hadronic parameters can be
eliminated after using SU(3), leaving 13. Considering $B\to K\bar K$ adds two
additional parameters. Note that dropping $C_{7,8}$ makes it impossible to fit
for new physics in these coefficients. In our SCET analysis all contributions
$C_{7-10}$ are included without needing additional hadronic parameters.

Finally, some analyses use additional ``dynamical assumptions'' and drop certain
combinations of reduced matrix elements in SU(3).  For example, the number of
parameters is often reduced by neglecting parameters corresponding to the so
called annihilation and exchange contributions.

\subsection{General parameterization of the amplitudes using SU(2)}
\label{sec:parametrization}
Using the SU(2) flavor symmetry, the most general amplitude parameterization for
the decay $B \to \pi \pi$ is
\begin{align}\label{pipigeneral}
A(\bar B^0 \to \pi^+ \pi^-) &= -\lambda_u^{(d)}   T_{\pi\pi} - \lambda_c^{(d)} 
    P_{\pi\pi}\\
A(\bar B^0 \to \pi^0 \pi^0) &= \! - \lambda_u^{(d)} C_{\pi\pi} -\! \lambda_c^{(d)}  
  ({EW}^T_{\pi\pi} - P_{\pi\pi})   \nn\\
\sqrt{2} A(B^- \to \pi^- \pi^0) &= -\lambda_u^{(d)}  (  T _{\pi\pi}+
    C_{\pi\pi}) - \lambda_c^{(d)}   {EW}^T_{\pi\pi}\nn
\end{align}
where we have used the unitarity of the CKM matrix
$\lambda_t^{(f)}=-\lambda_u^{(f)}-\lambda_c^{(f)}$. The amplitude parameter $
{EW}^T_{\pi\pi}$ receives contributions only through the electroweak penguin
operators $O_{7-10}$.  For $B \to K \pi$ decays we write
\begin{eqnarray}\label{Kpigeneral}
A(B^-\to \pi^- \bar K^0) &=& \lambda_u^{(s)}   A_{K \pi} +
\lambda_c^{(s)}   P_{K \pi} \\
\sqrt2 A(B^-\to \pi^0 K^-) &=&
-\lambda_u^{(s)} (  C_{K \pi} +   T_{K \pi} +   A_{K \pi})\nn\\ 
&& \hspace{-1cm}-
\lambda_c^{(s)}   (P_{K \pi} + {EW}^T_{K \pi}) \nn \\
A(\bar B^0\to \pi^+ K^-) &=&
-\lambda_u^{(s)}   T_{K \pi}\nn\\
&& \hspace{-1cm} - \lambda_c^{(s)}   (P_{K \pi} + {EW}^C_{K \pi}) \nonumber\\
\sqrt2 A(\bar B^0\to \pi^0 \bar K^0) &=&
- \lambda_u^{(s)}   C_{K \pi}  \nn\\
&& \hspace{-1cm} + \lambda_c^{(s)}  (P_{K \pi} -
 {EW}^T_{K \pi}+ {EW}^C_{K \pi})
\nonumber
\end{eqnarray}
Finally for $B \to K \bar K$ decays there is no SU(2) relation between the
amplitudes and we define
\begin{eqnarray}\label{KKgeneral}
A(B^-\to K^- K^0) &=& \lambda_u^{(d)} A_{KK} + \lambda_c^{(d)} P_{KK}\nn \\
A(\bar B^0\to K^0 \bar K^0) &=& \lambda_u^{(d)} B_{KK}  \nn\\
 && \hspace{-1cm}  + \lambda_c^{(d)} (P_{KK}+PA_{KK}+EW_{KK})\nn \\
A(\bar B^0\to K^- K^+) &= & \lambda_u^{(d)} E_{KK} - \lambda_c^{(d)} PA_{KK} 
 \,.
\end{eqnarray}

As mentioned before, after eliminating $\lambda_t^{(f)}$ there are four complex
hadronic parameters for $B \to \pi \pi$ and six for $B \to K \pi$.  The
additional relation one obtains in the limit $C_{7,8} \to 0$ is
\begin{align}
\label{EWTrelation}
 {EW}^T_{\pi\pi} &= \frac{3}{2} \frac{C_9+C_{10}}{C_1+C_2} ( {T}_{\pi\pi}
  +   C_{\pi\pi})\,,
\end{align}
where we have neglected terms quadratic in $C_9$ or $C_{10}$. 

The $EW$ amplitudes are purely from electroweak penguins, however there are also
electroweak penguin contributions in the other amplitudes as discussed further
in section~\ref{sectEW}.  Also, the hadronic parameters in
Eqs.~(\ref{pipigeneral})-(\ref{KKgeneral}) are a minimal basis of isospin
amplitudes, {\em not} graphical amplitude parameters.  In the appendix we show
how these amplitude parameters are related to the graphical amplitudes discussed
in~\cite{graphical1,graphical2,graphical3}.

\subsection{Additional relations in the SU(3) limit}
\label{sec:SU3relations}

In the limit of exact SU(3) flavor symmetry the parameters in the $\pi\pi$
system and the $K \pi$ system satisfy the two simple
relations~\cite{graphical1,Zeppenfeld,GrinsteinLebed}
\begin{eqnarray}
\label{SU3relation}
  T_{\pi \pi}+   C_{\pi \pi} =   T_{K \pi}+   C_{K \pi} \nn\\
 {EW}^T_{\pi\pi} =  {EW}^T_{K\pi} \,.
\end{eqnarray}
Thus, the hadronic parameters in the combined $K \pi$, $\pi \pi$ system can be
described by 8 complex parameters (15 real parameters after removing an overall phase), if no additional assumptions are made. 
A
choice for these parameters is
\begin{eqnarray}
  T_{\pi\pi}\,,\quad   C_{\pi\pi}\,,\quad   P_{\pi\pi}\,,\quad   A_{K \pi}\,,\nonumber\\
 {EW}^T_{K \pi}\,,\quad  {EW}^C_{K \pi}, \quad \Delta   C, \quad \Delta   P
\end{eqnarray}
where we have defined 
\begin{eqnarray}
\Delta   C &\equiv&   C_{K \pi} -    C_{\pi \pi} \nn\\
\Delta   P &\equiv&   P_{\pi \pi} -    P_{K \pi}
\end{eqnarray}
This can also be seen by relating these amplitude parameters directly to reduced
matrix elements in SU(3), which can be done with the help of the results in
Ref.~\cite{GrinsteinLebed}. As before, if the small Wilson coefficients $C_7$
and $C_8$ are neglected, we can again use the relation in
Eq.~(\ref{EWTrelation}) to eliminate one of the 8 complex hadronic parameters.

Four additional relations exist if the amplitudes for $B \to KK$ are included
\begin{eqnarray} \label{KKsu3}
A_{KK}  &=& A_{K\pi} \\
P_{KK}  &=& P_{K\pi} \nn \\
E_{KK}  &=& T_{K\pi} \!-\!  T_{\pi\pi} = - \Delta C \nn \\
PA_{KK}  &=& P_{\pi\pi}  \!-\! P_{K\pi} \!-\!  EW^C_{K\pi} 
 = \Delta P - EW^C_{K\pi} \nn \,.
\end{eqnarray}
In the limit of vanishing Wilson coefficients $C_7$ and $C_8$ there are two additional relation~\cite{graphical2}
\begin{eqnarray}
EW^C_{K \pi} &=& \frac{3}{4} \left[ \frac{C_9-C_{10}}{C_1-C_2} \left( A_{K\pi} - T_{\pi\pi} + C_{K \pi} + B_{KK}\right)\right.\nn\\
&&\hspace{-1cm}\left.-  \frac{C_9+C_{10}}{C_1+C_2} \left( A_{K\pi} - T_{\pi\pi} + C_{K \pi} -2C_{\pi\pi}- B_{KK}\right)\right]\nn\\
EW_{KK} &=& \frac{3}{2} \frac{C_9+C_{10}}{C_1+C_2} \left( B_{KK} - A_{KK} + E_{KK} \right)ñ
\end{eqnarray}

\subsection{Sum-Rules in $B\to K\pi$}
\label{sec:Sum-Rules}
In this section we review the derivation of two sum-rules for $B\to K\pi$, the
Lipkin sum-rule~\cite{Lipkin,GR2,soni} and CP
sum-rule~\cite{CPsum}. Higher order terms are kept and
will be used later on in assessing the size of hadronic corrections to these sum
rules using factorization. To begin we rewrite the SU(2) parameterization of the
amplitudes as
\begin{align}
 A(B^- \to \pi^- \bar K^0) \\
  &\hspace{-1.4cm} = \lambda_c^{(s)} P_{K\pi} \Big[ 1 -
 \frac{1}{2} \epsilon_A e^{-i\gamma} e^{i\phi_A} \Big]\,, \nn \\
 A(\bar B^0 \to \pi^+K^-) & \nn\\
 &\hspace{-1.4cm} = - \lambda_c^{(s)} P_{K\pi} \Big[ 1 \!+\!
 \frac{1}{2}\big(
  \epsilon_C^{ew} e^{i\phi_C^{ew}} \!-\! \epsilon_T e^{i\phi_T-i\gamma}\big) 
  \Big], \nn \\
 \sqrt{2} A(B^- \to \pi^0K^-) \nn\\
  &\hspace{-1.4cm} = - \lambda_c^{(s)} P_{K\pi} \Big[ 1\!+\!
 \frac{1}{2} \big( \epsilon_{T}^{ew} e^{i\phi_T^{ew}} \!-\! \epsilon
 e^{i\phi-i\gamma} 
  \big)  \Big], \nn \\
 \sqrt{2} A(\bar B^0 \to \pi^0\bar K^0) \nn\\
  &\hspace{-1.4cm} = \lambda_c^{(s)} P_{K\pi} \Big[ 1\!-\! \frac{1}{2} \big(
   \epsilon_{ew} e^{i\phi_{ew}}\!-\! \epsilon_C e^{i\phi_C-i\gamma} \big) 
  \Big], \nn 
\end{align}
where
\begin{align}
 \frac12\, \epsilon_T e^{i\phi_T} &=
     \bigg| \frac{\lambda_u^{(s)}}{\lambda_c^{(s)}} \bigg| \
     \frac{(-T_{K\pi})}{P_{K\pi}}  \,, \\
 \ \   \frac12\, \epsilon_C e^{i\phi_C} 
   &= \Big| \frac{\lambda_u^{(s)}}{\lambda_c^{(s)}} \Big| \
     \frac{(-C_{K\pi})}{P_{K\pi}} \,, \nn \\
 \ \   \frac12\, \epsilon_A e^{i\phi_A} 
   &= \Big| \frac{\lambda_u^{(s)}}{\lambda_c^{(s)}} \Big| \
     \frac{(-A_{K\pi})}{P_{K\pi}} \,, 
  \nn\\
  \frac12 \, \epsilon\: e^{i\phi} &= 
  \Big| \frac{\lambda_u^{(s)}}{\lambda_c^{(s)}} \Big| \
     \frac{(-T_{K\pi}- C_{K\pi}-A_{K\pi})}{P_{K\pi}}  \,,
  \nn
\end{align}
and
\begin{align}
  \frac12\, \epsilon_T^{ew}\: e^{i\phi_T^{ew}} &= 
    \frac{{EW}_{K\pi}^T}{P_{K\pi}}   \,,
  \\
  \frac12\, \epsilon_C^{ew} \: e^{i\phi_C^{ew}} &=   \
    \frac{{EW}_{K\pi}^C}{P_{K\pi}}   \,,
  \nn\\
  \frac12\, \epsilon_{ew} \: e^{i\phi_{ew}}&= 
     \frac{{EW}_{K\pi}^T-{EW}_{K\pi}^C}{P_{K\pi}}   \,.\nn
\end{align}
These parameters satisfy
\begin{align} \label{relation}
  \epsilon\: e^{i\phi} &= \epsilon_T\, e^{i\phi_T} + \epsilon_C\, e^{i\phi_C}
         + \epsilon_A\, e^{i\phi_A} \\
  \epsilon_{ew} e^{i\phi_{ew}} &= \epsilon^{ew}_T e^{i\phi^{ew}_T} -
         \epsilon^{ew}_C e^{i\phi^{ew}_C} \nn\,.
\end{align}
The non-electroweak $\epsilon$-parameters are suppressed by the small ratio of
CKM factors $|\lambda_u^{(s)}/\lambda_c^{(s)}| \simeq 0.024$ but are then
enhanced by a factor of $\sim 4$--$15$ by the ratio of hadronic amplitudes.  The
electroweak $\epsilon$-parameters are simply suppressed by their small Wilson
coefficients and end up being similar in size to the non-electroweak
$\epsilon$'s.

Next we define deviation parameters for the branching ratios
\begin{align}
  R_1 &= \frac{2 {\rm Br}(B^-\to \pi^0K^-)}
     {{\rm Br}(B^- \to \pi^- \bar K^0)} -1 
     \,,\\
    R_2 &= \frac{ {\rm Br}(\bar B^0\to \pi^-K^+)\tau_{B^-}}
     {{\rm Br}(B^- \to \pi^- \bar K^0)\tau_{B^0}} -1
     \,,\nn\\
    R_3 &= \frac{2 {\rm Br}(\bar B^0\to \pi^0\bar K^0)\tau_{B^-}}
     {{\rm Br}(B^- \to \pi^- \bar K^0)\tau_{B^0}} -1
    \,,\nn
\end{align}
and also rescaled asymmetries
\begin{align}
  &\Delta_1 = (1+R_1) A_{\rm CP}(\pi^0 K^-)
     \,,\\
  &\Delta_2 = (1+R_2)A_{\rm CP}(\pi^- K^+)
     \,,\nn\\
  &\Delta_3 = (1+ R_3) A_{\rm CP}(\pi^0 \bar K^0)  
    \,,\nn \\
  &\Delta_4 = A_{\rm CP}(\pi^- \bar K^0) \,. \nn
\end{align}
The division by ${\rm Br}(\pi^- \bar K^0)$ in the $\Delta_i$ asymmetries is not
necessary but we find it convenient for setting the normalization.  Expanding in
$\epsilon_A$ we find that to second order in the $\epsilon$ parameters the $R_i$
are
\begin{align}
  R_1 &=  \big[\epsilon_T^{ew}\cos\phi_T^{ew} \!-\! \epsilon \cos\phi
  \cos\gamma +\epsilon_A \cos\phi_A \cos\gamma  \big] \nn\\
  &+\! \Big[ \frac14\big( \epsilon^2 \!+\! \epsilon_T^{ew\,2}
   \!-\! \epsilon_A^{2} \big)
   \!-\! \frac{1}{2}\:
     \epsilon\,\epsilon_T^{ew} \cos\gamma \cos(\phi\!-\!\phi_T^{ew}) \nn\\
  &\ \ \ + (\epsilon_T^{ew}\cos\phi_T^{ew} \!-\! \epsilon \cos\phi
  \cos\gamma )\, \epsilon_A \cos\phi_A \cos\gamma \nn\\
  & \ \ \   + \epsilon_A^2 \cos^2\phi_A \cos^2\gamma 
   \Big] ,\nn\\
 R_2 &=  \big[\epsilon_C^{ew}\cos\phi_C^{ew} \!-\! \epsilon_T \cos\phi_T
  \cos\gamma \!+\! \epsilon_A \cos\phi_A \cos\gamma  \big] \nn\\
  &+\! \Big[ \frac14\big( \epsilon_T^2 \!+\! \epsilon_C^{ew\,2}
   \!-\! \epsilon_A^{2} \big)
   \!-\! \frac{1}{2}\:
     \epsilon_T\,\epsilon_C^{ew} \cos\gamma \cos(\phi_{C}^{ew}\!-\!\phi_T  )
   \nn\\
  &\ \ \ + (\epsilon_C^{ew}\cos\phi_C^{ew} \!-\! \epsilon _T\cos\phi_T
  \cos\gamma ) \, \epsilon_A \cos\phi_A \cos\gamma \nn\\
  & \ \ \ + \epsilon_A^2 \cos^2\phi_A \cos^2\gamma 
  \Big] ,\nn\\
 R_3 &=   \big[\!-\!\epsilon^{ew}\cos\phi_{ew} 
  \!+\! \epsilon_C \cos\phi_C \cos\gamma 
 \!+\!\epsilon_A \cos\phi_A \cos\gamma  \big] \nn\\
  &+\! \Big[ \frac14\big( \epsilon^{ew\,2} \!+\! \epsilon_C^{ew\,2}
   \!-\! \epsilon_A^{2} \big)
   \!-\! \frac{1}{2}\:
     \epsilon_C\,\epsilon^{ew} \cos\gamma \cos(\phi_{ew}\!-\!\phi_C) \nn\\
  &\ \ \  - (\epsilon^{ew}\cos\phi^{ew} \!-\! \epsilon_C\cos\phi_C
  \cos\gamma )\, \epsilon_A \cos\phi_A \cos\gamma \nn\\
  &\ \ \  + \epsilon_A^2 \cos^2\phi_A \cos^2\gamma 
   \Big] , 
\end{align}
and the $\Delta_i$ are
\begin{align}
  \Delta_1 &= \sin\gamma\big[ -\epsilon \sin\phi  
    -\frac12\,   \epsilon\, \epsilon_T^{ew} \sin(\phi\!-\!\phi_T^{ew}) \big]
    \nn\\
   & \ \ \ \times \big[ 1 + \epsilon_A \cos\phi_A \cos\gamma ]
    , \nn \\
 \Delta_2 &= \sin\gamma\big[-\epsilon_T \sin\phi_T
    -\frac12\,   \epsilon_T\, \epsilon_C^{ew} \sin(\phi_T\!-\!\phi_c^{ew}) 
   \big]   \nn\\
   & \ \ \ \times \big[ 1 + \epsilon_A \cos\phi_A \cos\gamma ]
  ,\nn\\
 \Delta_3 &= \sin\gamma\big[\epsilon_C \sin\phi_C 
    -\frac12\,   \epsilon_C\, \epsilon^{ew} \sin(\phi_C\!-\!\phi_{ew}) 
   \big]   \nn\\
   & \ \ \ \times \big[ 1 + \epsilon_A \cos\phi_A \cos\gamma ]
   ,\nn\\
 \Delta_4 &= \sin\gamma\big[-\epsilon_A \sin\phi_A \big]  
  \big[ 1 + \epsilon_A \cos\phi_A \cos\gamma ]\,. 
\end{align}
Note that these rescaled CP-Asymmetries are independent of the electroweak
penguin $\epsilon$'s at ${\cal O}(\epsilon)$.  This is not true for the original
asymmetries $A_{\rm CP}$.

Sum rules are derived by taking combinations of the $R_i$ and $\Delta_i$ which
cancel the ${\cal O}(\epsilon)$ terms.  The Lipkin sum rule is the statement
that
\begin{align} \label{Lipkinsum}
  & R_1 -R_2 + R_3 = {\cal O}(\epsilon^2) \nn \\ 
  & = \frac{1}{4} \big(\epsilon^2 - \epsilon_T^2 + \epsilon_C^2 -
    \epsilon_A^2 +\epsilon_T^{ew\,2} - \epsilon_C^{ew\,2} +\epsilon^{ew\,2} 
   \big) \nn\\
  & +\epsilon_A^2 \cos^2(\phi_A) \cos^2(\gamma) 
   -\frac{1}{2}\, \cos(\gamma) \big[  
   \epsilon_T^{ew} \epsilon \cos(\phi\!-\!\phi_T^{ew}) 
   \nn\\
  &
  -\epsilon_T \epsilon_C^{ew} \cos(\phi_T\!-\!\phi_C^{ew}) 
  + \epsilon_C \epsilon^{ew} \cos(\phi_C\!-\!\phi_{ew})  \big]  \,,
\end{align}
where we used the real part of Eq.~(\ref{relation}).  The CP-sum rule is the
statement that using the imaginary part of Eq.~(\ref{relation}) the ${\cal
  O}(\epsilon)$ terms cancel in the sum
\begin{align} \label{Deltasum}
 & \Delta_1 -\Delta_2 +\Delta_3 - \Delta_4  = {\cal O}(\epsilon^2) \nn \\
  &= -\frac{1}{2} \sin(\gamma) \big[ \epsilon_T^{ew}
 \epsilon \sin(\phi-\phi_T^{ew}) 
  - \epsilon_T \epsilon_C^{ew} \sin(\phi_T-\phi_C^{ew})
  \nn \\
 &\ \  
 -\epsilon_C \epsilon^{ew} \sin(\phi_C -\phi_{ew}) \big]  \,.
\end{align}
The accuracy of these sum rules can be improved if we can determine these ${\cal
  O}(\epsilon^2)$ terms using factorization. This is done in
section~\ref{sect_sumSCET}.

\section{Amplitude parameters in SCET} 
\label{sec:SCETtheory}

\subsection{General LO expressions}
\label{sec:LOexpressions}

The factorization of a generic amplitude describing the decay of a $B$ meson to
two light mesons, $B\to M_1 M_2$, has been analyzed using SCET~\cite{bprs}. Here
$M_1$ and $M_2$ are light (non-isosinglet) pseudoscalar or vector mesons. The
SCET analysis involves two stages of factorization, first between the scales
$\{m_b \mbox{ or } E_M\}^2\gg E_M\lqcd$, and second between $E_M\lqcd
\gg\lqcd^2$. Here we only consider the first stage of factorization where we
integrate out the scales $\{m_b,E_M\}$, and keep the most general
parameterization for physics at lower scales. It was shown in Ref.~\cite{bprs}
that a significant universality is already obtained after this first stage, in
particular there is only one jet function which also appears in semileptonic
decays to pseudoscalars and longitudinal vectors. This leads to the universality
of the function we call $\zeta_J^{BM}(z)$. We note that this also proves that
the second stage of matching is {\em identical} to that for the form factor, so
the SCET results for form factors in Refs.~\cite{ff} can immediately be applied
to nonleptonic decays if desired. A summary of the analysis of SCET operators
and matrix elements is given in Appendix~\ref{appSCET}.

After factorization at the scale $m_b$ the general LO amplitude for any $B\to M_1
M_2$ process can be written 
\begin{eqnarray}  \label{A0newfact}
A \!\!&=&\!\! 
\frac{G_F m_B^2}{\sqrt2}\! \bigg[ \bigg\{
   f_{M_1}\! \int_0^1\!\!\!\!du\, dz\,
    T_{1\!J}(u,z) \zeta^{BM_2}_{J}(z) \phi^{M_1}(u) 
   \nn \\
 &&\hspace{0.0cm}
   + f_{M_1} \zeta^{BM_2}\!\! \int_0^1\!\!\!\! du\, T_{1\zeta}(u) \phi^{M_1}(u)
  \bigg\} \!+\! \Big\{ 1\leftrightarrow 2\Big\} \nn\\
 && \hspace{0.0cm} + \lambda_c^{(f)} A_{c\bar c}^{M_1M_2} \bigg] ,  
\end{eqnarray}
where $\zeta^{BM}$ and $\zeta^{BM}_J$ are non-perturbative parameters describing
$B \to M$ transition matrix elements, and $A_{\rm c\bar c}^{M_1 M_2}$
parameterizes complex amplitudes from charm quark contractions for which
factorization has not been proven. Power counting implies $\zeta^{BM}\sim
\zeta^{BM}_J\sim (\Lambda/Q)^{3/2}$.  $T_{1\!J}(u,z)$ and $T_{1\zeta}(u)$ are
perturbatively calculable in an expansion in $\alpha_s(m_b)$ and depend upon the
process of interest.

It is useful to define dimensionless hatted amplitudes 
\begin{eqnarray}
 \hat A = \frac{A}{N_0} ({\rm GeV}^{-1})  \,,\quad 
  N_0 = \frac{G_F m_B^2}{\sqrt{2}} \,.
\end{eqnarray}
Using Eq.~(\ref{A0newfact}) we
find that the amplitude parameters in the $B \to \pi\pi$ system
are
\begin{eqnarray}
\label{pipiamps}
  \hat T_{\pi \pi} &=& - f_\pi \left[ \langle c_{1u} 
   +c_{4u}-c_{1t}^{ew}-c_{4t}\rangle_\pi \zeta^{B \pi} \right.\nn\\
&&\left. \hspace{0.38cm}
         + \langle (b_{1u} +b_{4u}-b_{1t}-b_{4t})
          \zeta_J^{B \pi}\rangle_\pi  \right]\nn\\
  \hat C_{\pi \pi} &=& - 
f_\pi \left[ \langle  c_{2u} -c_{2t}^{\rm ew}+c_{3t}^{\rm ew} +c_{4t}-c_{4u}
   \rangle_\pi \zeta^{B \pi} \right.\nn\\
&&\left. \hspace{0.38cm}
+               \langle (b_{2u} -b_{2t}^{\rm ew}+b_{3t}^{\rm ew} +b_{4t}-b_{4u}
 )\zeta_J^{B \pi} \rangle_\pi \right]\nn\\
  \hat P_{\pi \pi} &=&  -A_{cc}^{\pi \pi} 
  + f_\pi \left[ \langle c_{1t}^{\rm ew} + c_{4t}\rangle_\pi \zeta^{B \pi}
  \right.\nn\\
&&\left.\hspace{0.38cm} 
+  \langle (b_{1t}^{\rm ew} + b_{4t})\zeta_J^{B \pi} \rangle_\pi \right]\nn\\
 \hat {EW}^T_{\pi \pi} &=& 
f_\pi \left[ \langle c_{1t}^{\rm ew}+c_{2t}^{\rm ew}-c_{3t}^{\rm ew} \rangle_\pi \zeta^{B\pi}\right.\nn\\
&&\left. \hspace{0.38cm}
+ \langle (b_{1t}^{\rm ew}+b_{2t}^{\rm ew} - b_{3t}^{\rm ew})\zeta_J^{B\pi}\rangle_\pi  \right]\,.
\end{eqnarray}
For the $B \to  K\pi$ system we find
\begin{align}
\label{Kpiamps}
  \hat T_{K \pi} &= - f_K \left[ \langle c_{1u}
    -c_{1t}^{\rm ew}-c_{4t}+c_{4u}\rangle_{\bar K} \zeta^{B \pi} \right.\nn\\
&\left. \hspace{.38cm}
+ \langle (b_{1u} -b_{1t}^{\rm ew}-b_{4t}+b_{4u})\zeta_J^{B \pi}
\rangle_{\bar K} \right]\nn\\
  \hat C_{K \pi} &= - 
f_\pi \left[ \langle c_{2u} -c_{2t}^{\rm ew}+c_{3t}^{\rm ew} \rangle_\pi
  \zeta^{B\bar K} \right.\nn\\
&\left. \hspace{.38cm}
+               \langle (b_{2u} -b_{2t}^{\rm ew}+b_{3t}^{\rm ew}) \zeta_J^{B
  \bar K} \rangle_\pi \right]\nn\\
&\hspace{0.cm}
+ f_{K} \left[ \langle c_{4t}-c_{4u}\rangle_{\bar K} \zeta^{B \pi}+ \langle
  (b_{4t}-b_{4u}) \zeta_J^{B \pi} \rangle_{\bar K}\right]\nn\\
  \hat P_{K \pi} &= -A_{cc}^{K \pi} +f_K \left[ \langle
    c_{4t}\rangle_{\bar K} \zeta^{B \pi} 
+               \langle b_{4t}\zeta_J^{B \pi}\rangle_{\bar K}  \right]\nn\\
  \hat A_{K \pi} &= f_K \left[ \langle c_{4t}-c_{4u}\rangle_{\bar K} 
  \zeta^{B \pi}
+               \langle (b_{4t}-b_{4u}) \zeta_J^{B \pi} \rangle_{\bar K}\right]\nn\\
 \hat {EW}^T_{K \pi} &= 
f_K \left[ \langle c_{1t}^{\rm ew} \rangle_{\bar K} \zeta^{B\pi} 
+ \langle b_{1t}^{\rm ew} \zeta_J^{B\pi}  \rangle_{\bar K}\right]
\nn\\
&\hspace{0.7cm}
+f_\pi \left[ \langle c_{2t}^{\rm ew}-c_{3t}^{\rm ew} \rangle_\pi \zeta^{B \bar K}\right.\nn\\
&\left.\hspace{1.3cm}
+ \langle (b_{2t}^{\rm ew} - b_{3t}^{\rm ew}) \zeta_J^{B \bar K} \rangle_\pi \right]\nn\\
 \hat {EW}^C_{K \pi} &= f_K \left[ \langle c_{1t}^{\rm ew}\rangle_{\bar K} \zeta^{B \pi} 
+               \langle b_{1t}^{\rm ew}\zeta_J^{B \pi} \rangle_{\bar K} \right]\,.
\end{align}
Note that the dominant non-factorizable charm electroweak penguin contribution, is
absorbed into $A_{cc}$.  Finally, for the $B \to KK$ system we
find
\begin{align} \label{KKamps}
\hat A_{KK} &= \hat B_{KK}\\
 &= f_K \left[ \langle c_{4t}-c_{4u} \rangle_K \zeta^{B \bar K} 
+ \langle \left( b_{4t}-b_{4u}\right) \zeta^{B \bar K}_J \rangle_K \right]\nn\\
\hat P_{KK} &=  -A_{cc}^{K\bar K} 
 + f_K \left[ \langle c_{4t} \rangle_K \zeta^{B \bar K} 
+ \langle b_{4t} \zeta^{B \bar K}_J \rangle_{K} \right] \nn\\
 \hat E_{KK} &= \hat {PA}_{KK} = \hat {EW}_{KK} = 0 \,. \nn
\end{align}
In Eqs.~(\ref{pipiamps}-\ref{KKamps}) we have defined
\begin{eqnarray}
\langle c_i \rangle_M &=& \int_0^1 \!\!du \,\, c_i(u) \phi_M(u) \,, \\
\langle b_i\, \zeta_J^{BM_2} \rangle_{M_1} &=& \int_0^1 \!\!du\int_0^1 \!\! dz \,\, b_i(u,z)
    \phi_{M_1}(u)\zeta_J^{BM_2}(z)\,. \nn
\end{eqnarray}
We have also decomposed the Wilson coefficients of SCET operators defined 
in~\cite{bprs} as
\begin{eqnarray}
c_i^{(f)} &=& \lambda_u^{(f)} c_{iu} + \lambda_t^{(f)} c_{it} \nn\\
b_i^{(f)} &=& \lambda_u^{(f)} b_{iu} + \lambda_t^{(f)} b_{it}  
\end{eqnarray}
and in some equations we have split the contributions from strong ($O_{3-6}$)
and electroweak penguin operators ($O_{7-10}$) to the $c_{it}$ and
$b_{it}$
\begin{eqnarray}
c_{it} &=& c_{it}^p + c_{it}^{\rm ew} \,, \nn\\
b_{it} &=& b_{it}^p + b_{it}^{\rm ew} \,.
\end{eqnarray}
Note that to all orders in perturbation theory one has~\cite{graphical3}
\begin{eqnarray}
c_{1t}^p = c_{2t}^p = c_{3u} = c_{3t}^p  = 0\nn\\
b_{1t}^p = b_{2t}^p = b_{3u} = b_{3t}^p  = 0 \,.
\end{eqnarray}

All the hadronic information is contained in the SCET matrix elements
$\zeta^{B\pi}$, $\zeta_J^{B\pi}$, $\zeta^{BK}$, $\zeta_J^{BK}$,
$A_{cc}^{\pi\pi}$ and $A_{cc}^{K\pi}$, the decay constants $f_\pi$ and $f_K$,
and the light cone distribution functions of the light mesons $\phi_\pi(x)$,
$\phi_K(x)$, and $\phi_{\bar K}(x)$.

The Wilson coefficients $c_i$ and $b_i$ are insensitive to the long distance
dynamics and can therefore be calculated using QCD perturbation theory in terms
of the coefficients of the electroweak Hamiltonian, $C_i$.  Any physics beyond
the standard model which does not induce new operators in $H_W$ at $\mu=m_W$
will only modify the values of these Wilson coefficients, while keeping the
expressions for the amplitude parameters in Eqs.~(\ref{pipiamps}--\ref{KKamps})
the same.

We caution that although the amplitudes $\hat A_{K\pi}$ and $\hat A_{KK}$ do get
penguin contributions at this order, they will have subleading power
contributions from operators with large Wilson coefficients that can compete.
Therefore their leading order expressions presented here should not be used for
numerical predictions.  As mentioned earlier, the penguin amplitudes are kept to all 
orders in $\Lambda/E$. In all other amplitudes the power corrections are
expected to be genuinely down by $\Lambda/E$ when the hadronic parameters are of
generic size.  In the observables explored numerically below it will be valid
within our uncertainties to drop the small $\hat A_{K\pi}$ and $\hat A_{KK}$
amplitudes and so this point will not hinder us.

\subsection{SU(3) limit in SCET}
\label{sec:SU3SCET}

In the SU(3) limit the hadronic parameters for pions and kaons are equal.
This implies that
\begin{eqnarray}
&& \zeta^{B \pi} =\zeta^{B\bar K}\,, \qquad\quad
 \zeta_J^{B \pi} =\zeta_J^{B\bar K}\,,\nn\\[4pt]
 && \langle c_i \rangle_K = \langle c_i \rangle_\pi \,,
 \\[4pt]
 && \langle b_i \zeta_J^{B\bar K} \rangle_\pi = 
  \langle b_i \zeta_J^{B \pi} \rangle_K = \langle b_i \zeta_J^{B \pi}
  \rangle_\pi= \langle b_i \zeta_J^{B \bar K}
  \rangle_K
  \nn\,.
\end{eqnarray}
Furthermore
\begin{eqnarray} \label{Accsu3}
  A_{cc}^{K \pi} = A_{cc}^{\pi \pi} = A_{cc}^{K \bar K} \,.
\end{eqnarray}
To see this note that in SCET the light quark in the operator with two charm
quarks is collinear and can therefore not be connected initial $B$ meson without
further power suppression.  Without the use of SCET this so called ``penguin
annihilation'' contribution would spoil the relation in Eq.~(\ref{Accsu3}).

Using this we find two additional relations which are not true in a general SU(3)
analysis but are true in the combined SCET + SU(3) limit
\begin{eqnarray}
\Delta    C &=& C_{K\pi} - C_{\pi\pi} = 0 \,,\\
\Delta   P -   {EW}^C_{K \pi} &=&  P_{\pi\pi} - P_{K\pi}-  {EW}^C_{K \pi}
  =0   \,, \nn
\end{eqnarray}
where the zeroes on the RHS are ${\cal O}(m_s/\Lambda)+{\cal
  O}(\Lambda/E)$. Using the SU(3) relation in Eq.~(\ref{KKsu3}) we see that
these amplitudes are equal to ``exchange'' or ``penguin annihilation''
amplitudes that are power suppressed in SCET. 

\subsection{Results at LO in $\alpha_s(m_b)$}
\label{sec:treeresults}
While the $c_i$ are known at order $\alpha_s$, the $b_i$ are currently only
known at tree level. For consistency, we thus keep only the tree level
contributions to the $c_i$ as well. In this case they are independent of the
light cone fraction $u$ and thus $c_i(u) \equiv c_i$, and there occurs a single
nontrivial moment of the light-cone distribution function from the $b_i$ terms.
Since the parameter $A_{cc}^{M_1 M_2}\propto \alpha_s(2m_c)$ it would be
inconsistent to drop the $\alpha_s$ corrections in the penguin amplitudes.
However, as long as we have the free complex parameter $A_{cc}^{M_1 M_2}$ these
corrections are simply absorbed when we work with the full penguin amplitudes
$P_{M_1 M_2}$ using only isospin symmetry.  This is also true of chirally
enhanced power corrections in $P_{M_1 M_2}$.

Using LL values for the Wilson coefficients we find for the non-electroweak
amplitudes at $\mu=m_b=4.8\,{\rm GeV}$
\begin{eqnarray}\label{cexpr}
c_{1u} &=&
      C_1 \!+\! \frac{C_2}{N_c} = 1.025 \\
c_{2u} &=&
      C_2 \!+\! \frac{C_1}{N_c} = 0.121\nn\\
 c_{4t}^p &=& - \Big(C_4 + \frac{C_3}{N_c}\Big)+{\cal O}(C_1 \alpha_s)
  = 0.022 +{\cal O}(\alpha_s) \nn \,.
\end{eqnarray}
Here ${\cal O}(C_1\alpha_s ) $ indicate unsuppressed $\alpha_s$ corrections
that were computed in Ref.~\cite{QCDF} and verified in~\cite{pipiChay}. The contributions from the
operators $O_{7-10}$ give
\begin{eqnarray}
c_{1t}^{\rm ew} &=&
       - \frac32 \Big( C_{10} \!+\! \frac{C_9}{N_c} \Big) = 0.0021\nn\\
c_{2t}^{\rm ew}&=& - \frac32 \Big(C_9 \!+\! \frac{C_{10}}{N_c}\Big) = 0.0138
      \nn \\
c_{3t}^{\rm ew} &=&
       - \frac32  \Big( C_7 + \frac{C_8}{N_c}\Big) = -0.0010\nn\\
 c_{4t}^{\rm ew} &=&\frac{1}{2} \Big(C_{10} + \frac{C_9}{N_c}\Big) = -0.00068 
 \,.
\end{eqnarray}
 The coefficients $b_i(u,z)$ are
independent of the variable $z$ at leading order and we write $b_i(u,z) \equiv
b_i(u)$. For the non-electroweak amplitudes we have
\begin{eqnarray}\label{bexpr}
b_{1u}(u)&=&
      C_1 + \Big(1 + \frac{1}{\bar u} \Big)
      \frac{C_2}{N_c} \\
      &=& 1.025 -  \frac{ 0.249}{3\bar u}  \nn\\
  b_{2u}(u) &=&
      C_2 + \Big(1 + \frac{1}{\bar u} \Big)
      \frac{C_1}{N_c} \nn\\
      &=& 0.121 +  \frac{1.107}{3\bar u} \nn\\
  b_{4t}^p(u) &=& - C_4 - \Big(1 + \frac{1}{\bar u} \Big)\frac{C_3}{N_c} +{\cal O}(C_1\alpha_s )\nn\\
  &=& 0.022 -  \frac{0.011}{3\bar u} +{\cal O}(\alpha_s ) \nn
\end{eqnarray}
where ${\cal O}(C_1\alpha_s )$ denotes unknown  unsuppressed $\alpha_s$
corrections and $\bar u=1-u$. For the electroweak terms
\begin{eqnarray}
  b_{1t}^{\rm ew}(u) &=&
       - \frac32 \Big[C_{10} + 
    \Big(1 + \frac{1}{\bar u} \Big)
      \frac{C_9}{N_c} \Big] \nn\\
      &=& 0.0021 +\frac{0.0147}{3\bar u} \nn\\
b_{2t}^{\rm ew}(u)&=& - \frac32 \Big[ C_9 + \Big(1 + \frac{1}{\bar u} \Big)
      \frac{C_{10}}{N_c} \Big] \nn\\
      &=& 0.0138 - \frac{ 0.0029}{3\bar u}  \nn\\
b_{3t}^{\rm ew}(u) &=&
       - \frac32  \Big[ C_7 
     + \Big(1 - \frac{1}{\bar u} \Big)
      \frac{C_8}{N_c} \Big] \nn\\
      &=& -0.00010 +  \frac{0.00069}{3 u} \nn\\
b_{4t}^{\rm ew}(u) &=&\frac{1}{2} \Big[C_{10} +\Big(1 + \frac{1}{\bar u}
\Big)\frac{C_9}{N_c} \Big] \nn\\
&=& -0.00068 -  \frac{0.0049}{3\bar u}
 \,.
\end{eqnarray}
Note that only $b_{3t}^{\rm ew}$ involves $u$ and that this coefficient only
appears convoluted with pions in Eqs.~(\ref{pipiamps}-\ref{KKamps}). For pions
one can take $1/u\to 1/\bar u$ using charge conjugation and isospin.  Since the
$b$'s then only involve factors of $1/\bar u$ it is useful to define the
nonperturbative parameters
\begin{eqnarray}
 \beta_M = \int_0^1\!\!\!du\: \frac{\phi_M(u)}{3\bar u}  \,.
\end{eqnarray}

Using these values for the Wilson coefficients we obtain the amplitude
parameters in terms of the non-perturbative parameters in the $\pi \pi$ system
at $\mu=m_b$
\begin{eqnarray} \label{pipiparamscet}
\hat T_{\pi\pi} &=& -0.131 \big(\zeta^{B\pi} + \zeta_J^{B\pi} \big) +
  0.031\, \beta_\pi  \zeta_J^{B\pi}\nn\\
\hat C_{\pi\pi} &=& -0.017 \big( \zeta^{B\pi} + \zeta_J^{B\pi} \big)-
  0.144\, \beta_\pi  \zeta_J^{B\pi}\nn\\
\hat{EW}^T_{\pi\pi} &=& 0.0022 \big( \zeta^{B\pi}+\zeta_J^{B\pi} \big) +
 0.0015\, \beta_\pi\zeta_J^{B\pi} \nn\\
\hat P_{\pi\pi} &=& -A_{cc}^{\pi\pi} + 0.0030 \, \big( \zeta^{B\pi} +
   \zeta_J^{B\pi}\big) \nn\\
  && -0.0002\, \beta_\pi \zeta_J^{B\pi}
  + \Delta_{\pi\pi}^P \,,
\end{eqnarray}
where $\Delta_{\pi\pi}^P$ is the additional perturbative correction from
$a_{4t}$ and $b_{4t}$ at ${\cal O}(\alpha_s(m_b))$ which can involve larger
Wilson coefficients like $C_{1,2}$. (We could also include large power
corrections in $\Delta_{\pi\pi}^P$ assuming that such a subset could be uniquely
identified in a proper limit of QCD.) We do not need knowledge of
$\Delta_{\pi\pi}^P$ for our analysis since there are two unknowns in each of
$\hat P_{\pi\pi}$ and $A_{cc}^{\pi\pi}$ and we will simply fit for $\hat
P_{\pi\pi}$.

In the $K \pi$ system we find 
\begin{eqnarray}
\label{Kpiparamscet}
\hat T_{K\pi} &=& -0.160 \big(\zeta^{B\pi} + \zeta_J^{B\pi} \big)
 + 0.040 \,\beta_K  \zeta_J^{B\pi} \\
\hat C_{K\pi} &=& -0.003 \big(\zeta^{B\pi} + \zeta_J^{B\pi}\big)  
  + 0.003\, \beta_K  \zeta_J^{B\pi}\nn\\
&&-0.014 \big(\zeta^{B K} + \zeta_J^{B K}\big) 
  - 0.146\, \beta_\pi \zeta_J^{B K}\nn\\
\hat{EW}^T_{K\pi} &=&  0.0019 \big(\zeta^{B K} 
  + \zeta_J^{B K}\big) -0.0005 \beta_\pi
 \zeta_J^{B K} \nn\\
 && + 0.0003 \big( \zeta^{B\pi}+\zeta_J^{B\pi}\big) + 0.0023\, 
  \beta_K \zeta_J^{B\pi}\nn\\
\hat{EW}^C_{K\pi} &=& 
   0.0003 \big(\zeta^{B\pi} + \zeta_J^{B\pi}\big)+0.0023\, \beta_K \zeta_J^{B\pi}   \nn \\
\hat P_{K\pi} &=& -A_{cc}^{K\pi} 
   + 0.0034 \big( \zeta^{B\pi} + \zeta_J^{B\pi}\big)\nn\\
  && - 0.0026\, \beta_K  \zeta_J^{B\pi} + \Delta^P_{K\pi} \nn\\
\hat A_{K\pi} &=& 0.0034 \big( \zeta^{B\pi} + \zeta^{B\pi}_J ) -
  0.0026 \,\beta_K \zeta_J^{B\pi} +\Delta^A_{K\pi} \nn
\end{eqnarray}
where $\Delta_{K\pi}^P$ and $\Delta_{K\pi}^A$ are analogous corrections to
$\Delta_{\pi\pi}^P$.  For $P_{K\pi}$ the perturbative correction competes with
$A_{cc}^{K\pi}$. For $\hat A_{K \pi}$ power corrections could be in excess
of the leading order value, so that any numerical value for this amplitude is
completely unreliable at the order we are working.

Finally for the $B\to K\bar K$ amplitudes that have a contribution from the LO
factorization theorem  we have 
\begin{align}
\label{KKparamscet}
\hat A_{K K} &= 
   0.0034 (\zeta^{B K} + \zeta_J^{B K}) 
   - 0.0026 \beta_K \zeta_J^{B K}\nn\\
 & + {\cal O}(\alpha_s) \nn \\
\hat P_{K K}  &= - A_{cc}^{K\bar K} + 0.0034 (\zeta^{B K}+\zeta_J^{B
  K})\nn\\
 & - 0.0026 \beta_K \zeta_J^{B K}  + \Delta_{KK}^P
\end{align}
and $\hat B_{K K}=\hat A_{K K}$. Here the value of $\hat A_{KK}$ is not
reliable, since it will have large ${\cal O}(C_1 \Lambda/m_b)$ power corrections
that are likely to dominate.

\subsection{SCET Relations for EW penguin amplitudes} \label{sectEW}

Using the SCET results in the previous two sections it is simple to derive
relations that give the electroweak penguin contributions in terms of tree
amplitudes
\begin{align}
  \hat T_{M_1 M_2}^0 &= \hat T_{M_1 M_2}\big|_{C_{7-10}=0} \,,\nn\\
  \hat C_{M_1 M_2}^0 &= \hat C_{M_1 M_2}\big|_{C_{7-10}=0} \,.
\end{align}
Such relations are useful if one wishes to explore new physics scenarios that modify
the electroweak penguin parameters $C_{7-10}$ in $H_W$. To separate out all
electroweak penguin contributions in $B\to\pi\pi$ and using SCET together with isospin we 
define $\hat {EW}^C_{\pi\pi}$ by
\begin{align}
  & \hat T_{\pi\pi} = \hat T^0_{\pi\pi} \!+\! {\hat {EW}}_{\pi\pi}^C \,,
  &  \hat C_{\pi\pi} = \hat C^0_{\pi\pi} \!+\! {\hat {EW}}_{\pi\pi}^T
      \!-\! {\hat {EW}}_{\pi\pi}^C  \,, \nn\\
    &  \hat P_{\pi\pi} = \hat P^0_{\pi\pi} \!+\! {\hat {EW}}_{\pi\pi}^C \,.
\end{align}
At LO in SCET we find
\begin{align} \label{EW1}
  \hat {EW}_{\pi\pi}^C &= e_1\: \hat T_{\pi\pi}^0 + e_2 \: \hat C_{\pi\pi}^0\,,
     \ \nn \\ 
  \hat {EW}_{\pi\pi}^T &=e_3\: \hat T_{\pi\pi}^0 + e_4 \: \hat C_{\pi\pi}^0\,,
     \ 
\end{align}
where dropping $C_{3,4}$ relative to $C_{1,2}$ one finds
\begin{align}
  e_1 &= \frac{ C_{10} C_1 \!-\! C_9  C_2}
    {C_1^2\!-\! C_2^2} 
    = -2.9\!\times\! 10^{-4}  
   \,,\\
  e_2 &= \frac{ C_{9} C_1 \!-\! C_{10} C_2}
    {C_1^2\!-\! C_2^2}
    = -8.9\!\times\! 10^{-3}
    \,,\nn\\
  e_3 &= \frac{C_1(3C_{10}\!-\!3 C_7 \!-\! 2C_8 \!+\! 3 C_9) \!-\!
   3C_2(C_{10} \!+\!  C_8\!+\! C_9)}{2(C_1^2\!-\! C_2^2)} 
   \nn\\ 
   & = -1.5\times 10^{-2} 
  \,,\nn\\
  e_4 &= \frac{
   -C_2(3 C_{10} \!-\! 3C_7 \!-\! 2 C_8\!+\! 3C_9)\!+\!3C_1(C_{10} \!+\! C_8 \!+\!  C_9) }{2(C_1^2\!-\! C_2^2)}
        \nn \\
   & = -1.3\times 10^{-2} \,.\nn
\end{align}
The numbers quoted here are for the standard model LL coefficients. 

For $B\to K\pi$ we separate out the electroweak penguin contributions by writing
\begin{align}
  & \hat T_{K\pi} \!= \hat T^0_{K\pi} \!+\! \frac23 {\hat {EW}}_{\!K\pi}^C \,,
  \qquad  
  \hat A_{K\pi} = \hat P^0_{K\pi} \!-\! \frac13 {\hat {EW}}_{\! K\pi}^C \,,
   \nn \\
  &  \hat C_{K\pi}  = \hat C^0_{K\pi} \!+\! {\hat {EW}}_{\! K\pi}^T 
     \!-\! \frac23 {\hat {EW}}_{\! K\pi}^C \,,\nn\\
  &  \hat P_{K\pi}  = \hat P^0_{K\pi} \!-\! \frac13 {\hat {EW}}_{\! K\pi}^C \,,
\end{align}
and find that SCET+isospin gives
\begin{align}  \label{EW2}
  \hat {EW}_{K\pi}^C &= \frac{f_K}{f_\pi} 
   \Big( e_5\: \hat T_{\pi\pi}^0 + e_6 \: \hat C_{\pi\pi}^0 \Big)
      + e_{7} \: \hat T_{K\pi}^0\,,
    \ \nn
\end{align}
where dropping $C_{3,4}$ relative to $C_{1,2}$
\begin{align}
  e_5 &= -\frac{3 C_1(C_1 C_{9} \!-\! C_2 C_{10})}{2C_2 (C_1^2\!-\! C_2^2)}
      = -6.0 \times 10^{-2} \,,  \nn \\
  e_6 &= -\frac{3 (C_{10} C_2-  C_{9}C_1 )}{2(C_1^2\!-\! C_2^2)}
      = -1.3 \times 10^{-2} \,,  \nn \\
  e_7 &= \frac{3 C_9}{2 C_2} = 5.9\times 10^{-2} \,.
\end{align}
For the amplitude $EW^T_{K \pi}$ no such relation exists, if the inverse moments
$\beta_\pi$ and $\beta_K$ are taken as unknowns. One can still use the SU(3)
relation in Eq.~\ref{SU3relation} to equate $EW^T$ in the $K \pi$ and $\pi \pi$
system.

\subsection{Estimate of Uncertainties}
\label{sec:uncertainties}

\label{theoryerror}
These expressions of the amplitude parameters are correct at leading order in
$\lqcd/E_\pi$, and as we explained above, the complete set of Wilson
coefficients is currently only available at tree level. Thus, any amplitude
calculated from these SCET predictions has corrections at order
$\alpha_s(m_b)$ and $\lqcd/E_\pi$. Using simple arguments based on
dimensional analysis, we therefore expect corrections to any of these relations
at the 20\% level.  We are working to all orders in
$\alpha_s(\sqrt{\Lambda E})$, and so we avoid adding additional uncertainty from
expanding at this scale.

Note that we have allowed for a general amplitude $P_{\pi\pi}$, which
contributes to the reduced isospin matrix element $\langle {\bf 1/2} || {\bf 0}
|| {\bf 1/2} \rangle$ in the $K \pi$ system, and to the reduced isospin matrix
element $\langle {\bf 0} || {\bf 1/2} || {\bf 1/2} \rangle$ in the $\pi \pi$
system. All power correction contributing to the same reduced matrix element
will be absorbed into the value of the observable $P_{\pi\pi}$.  In the
following will thus fit directly for the parameters $P_{\pi\pi}$ and $P_{K\pi}$,
which reduces the theoretical uncertainties significantly. This implies that the
theoretical uncertainties on the amplitude parameters $\hat P_{\pi\pi}$, $\hat
P_{K \pi}$, and $\hat P_{K\bar K}$ are $\sim 3\%$ from isospin rather than $\sim
20\%$.  All other appreciable LO amplitude parameters are considered to have
uncertainties at the 20\% level.

Using this information, we can now estimate the size of corrections to the
individual observables.  For the decays $B \to \pi \pi$, contributions to the
total amplitude from $\hat P_{\pi\pi}$ and other amplitudes are comparable, such
that the whole amplitude receives ${\cal O}(20\%)$ corrections. This leads to
corrections to the branching ratios and CP asymmetries in $B \to \pi \pi$ of
order
\begin{eqnarray}
\Delta {\rm Br}( B \to \pi \pi) &\sim& {\cal O}(40\%)\nn\\
\Delta A_{\rm CP}( B \to \pi \pi) &\sim& {\cal O}(20\%)\,.
\end{eqnarray}
These large uncertainties can be avoided by relying on isospin to define most of
the parameters in the fit, as was done in Ref.~\cite{gammafit} in the
$\epsilon=0$ method for $\gamma$ which has significantly smaller theoretical
uncertainties.

For $B \to K \pi$ decays, the CKM factors and sizes of Wilson coefficients give
an enhancement of the amplitude parameter $\hat P_{K \pi}$ relative to the other
amplitude parameters by a factor of order 10.  Thus, the corrections to the
total decay rates are suppressed by a factor of 10, while corrections to CP
asymmetries, which require an interference between $\hat P_{K\pi}$ with other
amplitudes, remain the same. This gives
\begin{eqnarray}
\Delta {\rm Br}( B \to K \pi) &\sim& {\cal O}(5\%)\nn\\
\Delta A_{\rm CP}( B \to K \pi) &\sim& {\cal O}(20\%)\,.
\end{eqnarray}
One exception is the CP asymmetry in $B \to K^0 \pi^-$, which is strongly
suppressed due to the smallness of the parameter $\hat A_{K \pi}$. At subleading
order $\hat A_{K \pi}$ can receive corrections far in excess of the leading
order value, such that any numerical value of this CP asymmetry is completely
unreliable at the order we are working.  We include these estimates of power
corrections into all our discussions below.

\section{Implications of SCET}
\label{sec:SCETimplications}

There are several simple observations one can make from the LO SCET expressions
of the amplitude parameters
\begin{enumerate}
\item For $\zeta^{B M_i} \sim \zeta_J^{B M_i}$ one finds that $B\to M_1 M_2$
  decays naturally have $\hat C_{M_1 M_2} \sim \hat T_{M_1 M_2}$~\cite{bprs}, so
  there is no color suppression. If one instead takes $\zeta^{BM_i} \gg
  \zeta_J^{BM_i}$ as in Refs.~\cite{QCDF,PQCD} then the ``color suppressed''
  amplitude is indeed suppressed.
\item There is no relative phase between the amplitudes $C_{\pi\pi}$,
  $T_{\pi\pi}$, $T_{K\pi}$, $C_{K\pi}$, ${EW}^T_{K\pi}$ and ${EW}^C_{K\pi}$ and
  the sign and magnitude of these amplitudes can be predicted with SCET. This
  allows the uncertainty in the $K\pi$ sum-rules to be determined, as well as
  predictions for the relative signs of CP-asymmetries.
\item The contributions of electroweak penguins, $C_{7-10}$, can be computed
  without introducing additional hadronic parameters as discussed in
  Sec.~\ref{sectEW}.
\item The amplitude $\hat A_{K \pi}$ is suppressed either by $\Lambda/m_b$, by
  small coefficients $C_{3,4}$, or by $\alpha_s(m_b)$ compared with the
  larger $T_{K\pi}$ and $C_{K\pi}$ amplitudes
\item If one treats $\beta_K$ and $\beta_\pi$ as known, the amplitudes $\hat T_{K \pi}$ and $\hat{EW}^C_{K\pi}$
  are determined entirely through the hadronic parameters describing the $B \to
  \pi \pi$ system, implying that the branching ratios and CP-asymmetries for
  $B\to K^+\pi^-$ and $B^-\to \bar K^0 \pi^-$ only involve $2$ new parameters
  beyond $\pi\pi$.
\item In the combined SCET + SU(3) limit discussed in section~\ref{sec:SU3SCET}
  the parameters $P_{\pi\pi}\simeq P_{K\pi}\simeq P_{K\bar K}$, so we expect
  similar complex penguin amplitudes in $B\to K\pi$, $B\to \pi\pi$, and $B\to
  K\bar K$. \footnote{The analysis of ``chirally enhanced'' power 
  corrections in Ref.~\cite{QCDF} indicates that they will not break the
  equality in the SU(3) limit.}
\end{enumerate}
Using these observations allows us to make important predictions for the
observables, with and without performing fits to the data. Some of these have
already been discussed and we elaborate on the remaining ones below.

\subsection{The ratio $C/T$ and $EW^C/EW^T$}
\label{sec:C/T}

We first describe in more detail the first point in the above list. Most
literature has assumed that there is a hierarchy between the two amplitude
parameters $C_{M_1 M_2}$ and $T_{M_1 M_2}$, ie. that $\hat C_{M_1 M_2} \ll \hat
T_{M_1 M_2}$.  This assumption is based on the fact that in naive factorization
(in which $\zeta_J = 0$) one has $C_{M_1 M_2}/T_{M_1 M_2} \sim c_{2u}/c_{1u}\sim
0.1$.  The smallness of the ratio $c_{2u}/c_{1u}$ is due to the fact that the
dominant Wilson coefficient $C_1$ of the electroweak Hamiltonian is multiplied
by a factor of $1/N_c$ in $c_{2u}$, explaining the name ``color suppressed''
amplitude, plus additional accidental cancellations which reduce the value of
this ratio below 1/3.

In SCET, however, the Wilson coefficients $b_{1,2}$ contribute with equal
strength to the overall physical amplitude and can spoil the color
suppression~\cite{bprs}. In the $b_i$ terms for $C_{M_1 M_2}$ a factor of
$1/N_c=1/3$ occurs, however the hadronic parameter in the numerator is the
inverse moment of a light cone distribution function and is $\sim 3$.  Thus
numerically $\beta_{\pi,K} \simeq 1$, and setting $\beta_{\pi,K} = 1$ for
illustration we find
\begin{eqnarray}
\hat T_{\pi\pi} &=& -0.131 \zeta^{B\pi} -0.099\, \zeta_J^{B\pi}  \nn\\
\hat C_{\pi\pi} &=& -0.017 \zeta^{B\pi} -
  0.160\,  \zeta_J^{B\pi}\nn\\
  \hat T_{K\pi} &=& -0.160\, \zeta^{B\pi} -  0.120  \zeta_J^{B\pi}\nn\\
\hat C_{K\pi} &=& -0.003\, \zeta^{B\pi} -0.001\, \zeta^{B\pi}_J\nn\\
&&-0.014 \zeta^{BK} - 0.159\, \zeta_J^{BK}\,.
\end{eqnarray}
Thus, if $\zeta \sim \zeta_J$ it is easy to see that their is no ``color
suppression''.  On the other hand if $\zeta \gg \zeta_J$ as chosen in
Refs.~\cite{QCDF,PQCD} then one would have significant color suppression.

From Eqs.~(\ref{EW1}-\ref{EW2}) the size of the color-suppressed and color
allowed electroweak penguin amplitudes in $\pi\pi$ and $K\pi$ are directly
related to that of $C_{M_1 M_2}$ and $T_{M_1 M_2}$. Thus if $C_{M_1 M_2} \sim
T_{M_1 M_2}$ then SCET predicts that $EW^{C}_{M_1 M_2}\sim EW^T_{M_1 M_2}$.

\begin{table}[t!]
\begin{tabular}{cc|c}
 & Parameter & Measured value  \\
\hline
  & $m_{B}$ &  $(5279.4 \pm 0.5)$ MeV~\cite{pdg}\\
  & $\tau_{B^0}$ &  $(1.528 \pm 0.009)$ ps~\cite{HFAG}\\
  & $\tau_{B^+}$ & $(1.643 \pm 0.010)$ ps~\cite{HFAG}\\
  & $\beta$ & $0.379 \pm 0.022$~\cite{HFAG}\\
  & $f_\pi$ & $(130.7 \pm 0.4)$ MeV~\cite{pdg}\\
  & $f_K$ &  $(159.8 \pm 1.5)$ MeV~\cite{pdg}  \\
  & $|V_{ud}|$ & $0.9739 \pm 0.0003$~\cite{ckm05}\\
  & $|V_{us}|$ & $0.2248 \pm 0.0016$~\cite{ckm05}\\
  & $|V_{cd}|$ & $0.2261 \pm 0.0010$~\cite{CKMfitter}\\
  & $|V_{cs}|$ & $0.9732 \pm 0.0002$~\cite{CKMfitter}\\
  & $|V_{cb}|$ & $(41.6 \pm 0.5) \times 10^{-3}$~\cite{HFAG,globalfit}\\
  \hline
 & $|V_{ub}|^{\rm incl}$ & $(4.39 \pm 0.34) \times 10^{-3}$~\cite{HFAG,theoryVub} \\
 & $|V_{ub}|^{\rm excl}$ & $(3.92 \pm 0.52) \times 10^{-3}$~\cite{Lattice,agrs,LP} \\
 & $|V_{ub}|^{\rm global}_{\rm CKM}$ & 
    $(3.53 \pm 0.22) \times 10^{-3}$~\cite{CKMfitter} \\
 & $|V_{ub}|^{\rm here}$ & 
    $(4.25 \pm 0.34) \times 10^{-3}$ \\
  \hline
\end{tabular}
\caption{Summary of well measured input parameters. For our central value for
  $|V_{ub}|$ we use a weighted average of the inclusive~\cite{HFAG} and
  exclusive~\cite{LP} with a slightly inflated error.  Use $m_t=174.3\,{\rm GeV}$.
\label{tab3}}
\end{table}

\subsection{$B \to \pi \pi$ with Isospin and ${\rm Im}(C_{\pi\pi}/T_{\pi\pi})=0$}
\label{sec:pipi}

Using only SU(2) there are a total of 5 hadronic parameters describing the
decays $B \to \pi\pi$, in addition to a weak phase. The 6 measurements allow in
principle to determine all of these parameters as was first advocated by Gronau
and London~\cite{GL}. Unfortunately, the large uncertainties in the direct CP
asymmetry of $\bar B^0 \to \pi^0 \pi^0$ do not allow for a definitive analysis
at the present time (ie. it currently gives $65^\circ < \alpha <
200^\circ$~\cite{CKMfitter}). It was shown in Ref.~\cite{gammafit} that one can
use SCET to eliminate one of the 5 hadronic SU(2) parameters, since $\epsilon=
{\rm Im}(C_{\pi\pi}/T_{\pi\pi})\simeq 0$, and then directly fit for the
remaining four hadronic parameters and the weak angle $\gamma$, which
substantially reduces the uncertainty. Using the most recent data shown in
section~\ref{sec:introduction}, we find
\begin{eqnarray} \label{SCETgamma}
\gamma^{\pi\pi} = 83.0^\circ {}^{+7.2^\circ}_{-8.8^\circ} \pm 2^\circ
\end{eqnarray}
where the first error is from the experimental uncertainties, while the second
uncertainty is an estimate of the theoretical uncertainties from the expansions
in SCET, estimated by varying $\epsilon=\pm 0.2$ as explained in
Ref.~\cite{gammafit}. This value is in disagreement with the results from a
global fit to the unitarity triangle
\begin{align}
  \gamma_{\rm\, global}^{\rm\, CKMfitter} 
     &= 58.6^\circ {}^{+6.8^\circ}_{-5.9^\circ} 
  \,,\nn\\
  \gamma_{\rm\, global}^{\rm\, UTfit} 
     &= 57.9^\circ \pm 7.4^\circ \,,
\end{align}
at the $2$-$\sigma$ level. A more sophisticated statistical analysis can be
found in Ref.~\cite{GHLP}. The errors in Eq.~(\ref{SCETgamma}) are slightly
misleading because they do not remain Gaussian for larger $\epsilon$. At
$\epsilon=0.3$ the deviation drops to $1.5$-$\sigma$, and at $\epsilon=0.4$ it
drops to $0.5$-$\sigma$. The result in Eq.~(\ref{SCETgamma}) is  consistent
with the direct measurement of this angle which has larger errors~\cite{HFAG}
\begin{eqnarray}
\gamma^{DK} = 63^\circ {}^{+15^\circ}_{-12^\circ}\,. \qquad
\end{eqnarray}

\begin{figure}[t!]
\begin{center}
\includegraphics[width=9cm]{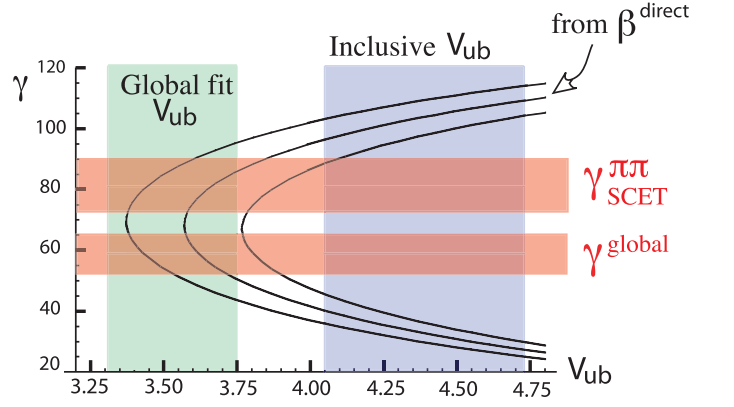}
\caption{Comparison of constraints on $V_{ub}$ and $\gamma$ from i) the direct
  measurement of $\beta$, ii) current HFAG value for inclusive $|V_{ub}|$, iii)
  global fit value of $\gamma$, iv) $|V_{ub}|$ as output from the global fit
  ~\cite{HFAG,CKMfitter}, and v) results for $\gamma$ from the small $\epsilon$
  analysis of $B\to\pi\pi$ decays~\cite{gammafit}. All errors bands are
  $1$-$\sigma$.}
\label{figgammaVub}
\end{center}
\end{figure}

It is interesting to note that the global fit for $\beta$ plus the inclusive
determination of $|V_{ub}|$ in table~\ref{tab3} also prefers larger values of
$\gamma$ as shown in Fig.~\ref{figgammaVub}.  It will be quite interesting to see
how these hints of discrepancies are sharpened or clarified in the future. In the
remainder of this paper, we will show results for $\gamma = 83^\circ$ and
$\gamma = 59^\circ$ to give the reader an indication of the $\gamma$ dependence
of our results.

The phase of the amplitude $A_{cc}^{\pi\pi}$ is mostly determined from the CP
asymmetries in $B \to \pi^+ \pi^-$. In particular, as can be seen from the
general parameterization of the amplitudes in Eq.~(\ref{pipigeneral}), the sign
of the direct CP asymmetry $C({\pi^+\pi^-})$ is correlated with the relative
sign between $\hat P_{\pi\pi}$ and $\hat T_{\pi\pi}$ and the sign of the
asymmetry $C({\pi^0\pi^0})$ with that between $\hat C_{\pi\pi}$ and $\hat
P_{\pi\pi}$. Since there is no relative phase between the amplitudes $\hat
C_{\pi\pi}$ and $\hat T_{\pi\pi}$ at LO in SCET, the sign of the direct CP
asymmetry in $B \to \pi^0\pi^0$ is thus expected to be positive based on the
negative experimental value for $C(\pi^+\pi^-)$~\cite{GHLP}. This expectation is in
disagreement with the direct measurement shown in Table~\ref{dataall}. Using the
values of the hadronic parameters from the previous fit we find for $\gamma =
83^\circ$
\begin{eqnarray}
C(\pi^0 \pi^0) = 0.49 \pm 0.12\pm 0.23\,,
\end{eqnarray}
while for $\gamma = 59^\circ$ we find
\begin{eqnarray}
C(\pi^0 \pi^0) = 0.61 \pm 0.19\pm 0.19\,.
\end{eqnarray}
These values are $1.7 \sigma$ from the measured value, if we add the theoretical
and experimental errors in quadrature.

\subsection{The decays $B \to \pi \pi$ in SCET}
\label{sec:pipi2}

For $\gamma = 83^\circ$ a fit of the four SCET parameters to the $B \to \pi \pi$
data excluding the direct CP asymmetry in $B \to \pi^0 \pi^0$ gives
\begin{align}
\label{pipiparams83}
\zeta^{B \pi} &= (0.088 \pm 0.019 \pm 0.045)\ \Big( \frac{4.25\times
  10^{-3}}{|V_{ub}|}\Big) \nn\\
\zeta_J^{B \pi} &= (0.085 \pm 0.016 \pm 0.031)\ \Big( \frac{4.25\times
  10^{-3}}{|V_{ub}|}\Big) \nn\\
10^3\hat P^{\pi\pi} &= (5.5 \pm 0.8\pm 1.3)e^{i(151\pm8\pm 6)^\circ},
\end{align}
while for $\gamma = 59^\circ$ we find
\begin{align}
\label{pipiparams59}
\zeta^{B \pi} &= (0.093 \pm 0.023 \pm 0.035)\ \Big( \frac{4.25\times
  10^{-3}}{|V_{ub}|}\Big)\nn\\
\zeta_J^{B \pi} &= (0.10 \pm 0.016 \pm .022)\ \Big( \frac{4.25\times
  10^{-3}}{|V_{ub}|}\Big)\nn\\
10^3 \hat P^{\pi\pi} &= (2.6 \pm 0.9\pm 0.8)e^{i(103\pm19\pm 16)^\circ}
  \! .
\end{align}
The first error is purely from the uncertainties in the experimental data, while
the second error comes from adding our estimate of the theory uncertainties
discussed in section~\ref{theoryerror}. For both values of $\gamma$ one finds
that $|P/T| \sim 0.25$. Note that this ratio of $P/T$ does not include the ratio
of the CKM factors. The ratio relevant for the decays is
$|\lambda_c^{(d)}/\lambda_u^{(d)}| |P/T| \sim 0.6$. 

It is interesting to compare the result for $\hat P^{\pi\pi}$ extracted from the
data with that from the purely perturbative penguin computed in
Ref.~\cite{QCDF} (in two scenarios for the input parameters),
\begin{align}
 10^3\hat P^{\rm default}_{\rm QCDF} &= -1.0 -[ 0.1 + 0.3 i+ \ldots] \\
  &\hspace{-0.6cm} 
  - (1.7 + 0.0 X_H+ .047 X_A^2) -\{0.3 + 0.1 i  \}\nn\\
 &= -3.1 -0.4i - 0.047 X_A^2 \,, \nn\\
  10^3\hat P^{\rm S2}_{\rm QCDF} &= -0.9 -[ 0.0 + 0.1 i+ \ldots] \nn\\
  &\hspace{-0.6cm} -(2.0 + 0.0 X_H+ .063 X_A^2) -\{0.3 + 0.3 i \} \nn \\
  &= -3.2 -0.4i - 0.063 X_A^2 \,, \nn .
\end{align}
Here the terms $[\ldots]$ are $\alpha_s(m_b)$ corrections, the terms $(\ldots)$
are chirally enhanced power corrections with parameters $X_H$ and $X_A$, and
$\{\ldots \}$ are perturbative corrections to these. We observe that the
magnitude of the perturbative $\pi\pi$ penguin is of similar size to that from
the data for $\gamma=59^\circ$, but has a small strong phase is in contrast to
the large strong phase seen in the data.

The correlation between $\zeta^{B \pi}$ and $\zeta_J^{B \pi}$ in
Eqs.~(\ref{pipiparams83},\ref{pipiparams59}) is about $-0.8$, so that the
heavy-to-light form factor, which is given by the sum of these two parameters is
determined with much smaller uncertainties than one would obtain by naively
adding the two individual errors in quadrature. For $\gamma = 83^\circ$ we find
\begin{eqnarray}
F^{B\pi}(0) = (0.17 \pm 0.01 \pm 0.03)\ \Big( \frac{4.25\times
  10^{-3}}{|V_{ub}|}\Big)
\end{eqnarray}
while for $\gamma = 59^\circ$ we find
\begin{eqnarray}
F^{B\pi}(0) = (0.19 \pm 0.01 \pm 0.03)\ \Big( \frac{4.25\times
  10^{-3}}{|V_{ub}|}\Big)
\end{eqnarray}

\subsection{The decays $B \to K \pi$}
\label{sec:Kpi}

For these decays the penguin amplitudes are enhanced by the ratio of CKM matrix
elements $|\lambda_c^{(s)}/\lambda_u^{(s)}| \sim 40$. Thus, the relevant ratio
of penguin to tree amplitudes is $|\lambda_c^{(s)}/\lambda_u^{(s)}| |P/T| \sim
10$ and the $B \to K \pi$ decays are penguin dominated. If one were to only keep
the penguin contributions to these decays the relative sizes of the branching
ratios would be determined by simple Clebsch-Gordon coefficients
\begin{align}
\label{clebschrelation}
 {\rm Br}(\pi^0 \bar K^0) \simeq {\rm Br}(\pi^0 K^-)\simeq
   \frac{{\rm Br}(\pi^+ K^-)}{2} \simeq  \frac{{\rm Br}(\pi^- \bar K^0)}{2} \,.
\end{align}
Deviations from this relation are determined at leading order in the power
counting by the non-perturbative parameters $\zeta^{BM}$ and $\zeta_J^{BM}$. 
To see how well the current data constrains deviations from this result we
can look at the following ratios of branching fractions
\begin{align} \label{R}
  R_1 &= \frac{2 {\rm Br}(B^-\to \pi^0K^-)}{{\rm Br}(B^- \to \pi^- \bar K^0)} -1 
    = 0.004 \pm 0.086  \,,\\
    R_2 &= \frac{ {\rm Br}(\bar B^0\to \pi^-K^+)\tau_{B^-}}{{\rm Br}(B^- \to \pi^-
      \bar K^0)\tau_{B^0}} -1
    = -0.157 \pm 0.055 \,,\nn\\
    R_3 &= \frac{2 {\rm Br}(\bar B^-\to \pi^0\bar K^0)\tau_{B^-}}{{\rm Br}(\bar B^0 \to
      \pi^- \bar K^0)\tau_{B^0}} -1
    = 0.026 \pm 0.105\,,\nn
\end{align}
and the rescaled asymmetries
\begin{align} \label{rAsym}
  &\Delta_1 = (1+R_1) A_{\rm CP}(\pi^0 K^-)
    = 0.040 \pm 0.040  \,,\\
  &\Delta_2 = (1+R_2)A_{\rm CP}(\pi^- K^+)
    = -0.097 \pm 0.016 \,,\nn\\
  &\Delta_3 = (1+ R_3) A_{\rm CP}(\pi^0 \bar K^0)  
    = - 0.021 \pm 0.133\,,\nn \\
  &\Delta_4 = A_{\rm CP}(\pi^- \bar K^0) = -0.02\pm0.04 \,.
\end{align}

These ratios have been defined by normalizing each branching ratio to the decay
$B^- \to \pi^0 \bar K^0$. If we drop the small amplitude parameter $A_{K
  \pi}$ then this channel measures the penguin,
\begin{eqnarray}
A(B^- \to \pi^- \bar K^0) = \lambda_c^{(s)} P_{K \pi}\,,
\end{eqnarray}
and the direct CP asymmetry is expected to be small. 

A simple test for the consistency of the $K\pi$ data is given by the Lipkin
sum-rule for branching ratios~\cite{Lipkin}, and a sum-rule for the
CP-asymmetries~\cite{CPsum}
\begin{eqnarray} \label{sum0}
  R_1 - R_2 +R_3 = 0 \,,\quad
  \Delta_1 - \Delta_2 + \Delta_3  -\Delta_4 = 0 \,, 
\end{eqnarray}
which are both second order in the ratio of small to large amplitudes as
discussed in section~\ref{sec:Sum-Rules}. The current data gives
\begin{eqnarray}
 && R_1 - R_2 +R_3 = (0.19 \pm 0.15)^{\rm expt} \,,\nn\\
 && \Delta_1 - \Delta_2 + \Delta_3 -\Delta_4 = (0.14 \pm 0.15)^{\rm expt} \,.
\end{eqnarray}
Thus, so far this global test does not show a deviation from the expectation.

SCET provides us with additional tests for the $K\pi$ data.  It turns out
that the current data is not precise enough to determine the values of
$\zeta^{BK}$ and $\zeta_J^{BK}$. These two parameters only contribute 
to the two decays $B^- \to \pi^0 K^-$ and $\bar B^0 \to \pi^0 \bar K^0$, which 
have neutral pions and larger experimental uncertainties. As we will explain, the data on these
decays seems to favor a negative value of $\zeta_J^{BK}$, but that would imply a negative value for $\lambda_B$, the first inverse moment of the $B$ meson wave function, contrary to any theoretical prejudice. One can use the fact that the only
sizeable strong phase is in the value of the parameter $A_{cc}^{K \pi}$ to
determine the predicted size of the deviations from the above relations and also
the signs and hierarchy for the CP asymmetries.

\subsubsection{Sum-Rules in $B\to K\pi$}
\label{sect_sumSCET}

In SCET positive values of $\zeta^{BM}$ and $\zeta_J^{BM}$ imply that the phase
of $-T_{K\pi}$, $-C_{K\pi}$, ${EW}_{K\pi}^{C,T}$, and
${EW}_{K\pi}^{T}-{EW}_{K\pi}^{C}$ are the same. This can be seen from
Eqs.(\ref{pipiparamscet}-\ref{Kpiparamscet}). Therefore this implies that these
amplitudes have a common strong phase $\delta$ relative to the penguin $\hat
P_{K\pi}$. Using the notation and results from section~\ref{sec:Sum-Rules} we
have
\begin{align}
  \phi_T = \phi_C = \phi = \phi_T^{ew} = \phi_C^{ew} = \phi_{ew} = \delta 
  \,.
\end{align}
At LO in SCET one can drop the $A_{K\pi}$ amplitude ($\epsilon_A=0$) and write
\begin{align}
 A(\bar B^0 \to \pi^- \bar K^0) &= \lambda_c^{(s)} P_{K\pi} \,,\\
 A(\bar B^0 \to \pi^+K^-) &= - \lambda_c^{(s)} P_{K\pi} \Big[ 1 \!+\!
 \frac{e^{i\delta}}{2} \big( \epsilon_C^{ew} \!-\! \epsilon_T e^{-i\gamma} \big) 
  \Big], \nn \\
 \sqrt{2} A(B^- \to \pi^0K^-) &= - \lambda_c^{(s)} P_{K\pi} \Big[ 1\!+\!
 \frac{e^{i\delta}}{2} \big( \epsilon_{T}^{ew} \!-\! \epsilon e^{-i\gamma} \big) 
  \Big], \nn \\
 \sqrt{2} A(\bar B^0 \to \pi^0\bar K^0) &= \lambda_c^{(s)} P_{K\pi} \Big[ 1\!-\!
 \frac{e^{i\delta}}{2} \big( \epsilon^{ew} \!-\! \epsilon_C e^{-i\gamma} \big) 
  \Big], \nn 
\end{align}
where the $\epsilon$-parameters are all positive and satisfy
\begin{eqnarray}
  \epsilon = \epsilon_T + \epsilon_C \,, \quad
  \epsilon^{ew} = \epsilon_T^{ew} - \epsilon_C^{ew} \,,
\end{eqnarray}
and 
\begin{eqnarray} \label{epsineq}
  \epsilon>\epsilon_C \,,\quad 
   \epsilon>\epsilon_T \,,\quad
  \epsilon_T^{ew} > \epsilon_C^{ew}\,,\quad
  \epsilon_T^{ew} > \epsilon^{ew} .
\end{eqnarray}

From the decomposition in terms of SCET parameters we can determine the
magnitudes of the $\epsilon$-parameters in terms of the $\zeta$'s. The rate
${\rm Br}(B^-\to \pi^- \bar K^0)$ determines 
\begin{align}
 10^3 \, |\hat P_{K\pi}|\simeq 5.5 \pm 0.1 \pm 0.1
\end{align}
and using 
\begin{align}
  \bigg| \frac{\lambda_u^{(s)}}{\lambda_c^{(s)}} \bigg| = 0.0236 \,,
\end{align}
we find
\begin{eqnarray}
   \epsilon_T &\simeq& 1.40 (\zeta^{B\pi} \!+\!\zeta_J^{B\pi}) 
    + 0.35 \beta_{\bar K} \zeta_J^{B\pi} \,,\nn\\
   \epsilon_C &\simeq& 0.12 (\zeta^{B K} \!+\!\zeta_J^{B K}) 
    + 1.27 \beta_\pi \zeta_J^{B\bar K} \,,\nn\\
    && +0.03(\zeta^{B \pi} \!+\!\zeta_J^{B \pi}) 
    - 0.02 \beta_{\bar K} \zeta_J^{B\pi} \nn\\
  \epsilon_T^{ew} &\simeq& 0.71 (\zeta^{B K} \!+\!\zeta_J^{B K}) 
    -0.17 \beta_{\pi} \zeta_J^{BK}\nn\\
    && + 0.12 (\zeta^{B \pi} \!+\!\zeta_J^{B \pi}) + 0.87 \beta_K \zeta_J^{B
      \pi}  \,, \nn\\ 
   \epsilon_C^{\rm ew} &\simeq& 0.12 (\zeta^{B\pi} \!+\!\zeta_J^{B\pi}) 
    + 0.87 \beta_\pi \zeta_J^{B\bar K} \,.
\end{eqnarray}
Generically $\zeta^{BM}+\zeta^{BM}_J\sim 0.15-0.25$ and $\zeta^{BM}_J\sim
0.05-0.15$ so that $\epsilon_T$, $\epsilon_C$, $\epsilon_T^{ew}$,
$\epsilon_C^{ew}$ are $\sim 0.1$--$0.4$ and can be thought of as expansion
parameters.

To estimate the SM deviations from the results in Eq.~(\ref{sum0}) we take the
${\cal O}(\epsilon^2)$ terms in Eqs.~(\ref{Lipkinsum},\ref{Deltasum}) and
independently vary the parameters in the conservative ranges
$\zeta^{B\pi}+\zeta^{B\pi}_J= 0.2\pm0.1$, $\beta_{\bar K}\zeta_J^{B\pi} = 0.10
\pm 0.05$, $\zeta^{B\bar K}+\zeta^{B\bar K}_J= 0.2\pm0.1$, $\beta_\pi
\zeta_J^{B\bar K} = 0.10 \pm 0.05$, $\epsilon_A = 0 \pm 0.1$, $\gamma=70^\circ
\pm 15^\circ$, arbitrary $\phi_A$ and all phase differences $\Delta \phi =
0^\circ \pm 30^\circ$.  For the Lipkin sum rule this gives
\begin{eqnarray}\label{sumerrors1}
  R_1 - R_2 + R_3 = 0.028 \pm 0.021 \,,
\end{eqnarray}
and for the CP-sum rule
\begin{eqnarray}\label{sumerrors2}
 \Delta_1 -\Delta_2 + \Delta_3 -\Delta_4 = 0 \pm 0.013 \,.
\end{eqnarray}
Experimental deviations that are larger than these would be a signal for new
physics. The CP-sum rule has significantly smaller uncertainty than the Lipkin
sum-rule. This can be understood from the expression
\begin{align}
 & \Delta_1 -\Delta_2 +\Delta_3  - \Delta_4 
  = -\frac{1}{2} \sin(\gamma) \big[ \epsilon_T^{ew}
 \epsilon \sin(\phi-\phi_T^{ew})\nn \\
 &\ \   - \epsilon_T \epsilon_C^{ew} \sin(\phi_T-\phi_C^{ew})
 -\epsilon_C \epsilon^{ew} \sin(\phi_C -\phi_{ew}) \big] \nn\\ 
 & \times \big(1+ {\cal O}(\epsilon_A)\big)  \,.
\end{align}
All terms involve one of the smaller electroweak penguin $\epsilon$-parameters,
and in SCET all the phase differences are small, both of which give a further
suppression over the Lipkin sum-rule.  Since the CP sum-rule is always
suppressed by at least three small parameters it is likely to be very accurate.

\subsubsection{CP-Asymmetry Sign Correlations} \label{sec:CPsign}

For the asymmetry parameters up to smaller terms of ${\cal O}(\epsilon^2)$ we
have
\begin{align}
  \Delta_1 = -\epsilon \sin(\delta) \sin(\gamma) ,\nn\\
 \Delta_2 = - \epsilon_T \sin(\delta)\sin(\gamma) ,\nn\\
 \Delta_3 = \epsilon_C \sin(\delta) \sin(\gamma)  .
\end{align}
Thus,  we immediately have the following predictions
\begin{align}
 \mbox{i)  }\ \  &  \Delta_1, \Delta_2, -\Delta_3 \mbox{ have
the same sign} \,,\nn\\
 \mbox{ii)  }\ \  &    |\Delta_1| \gtrsim |\Delta_2| \,,\quad  
 |\Delta_1| \gtrsim |\Delta_3| \,,
\end{align}
where i) depends only on the fact that positive $\zeta$'s gives positive
$\epsilon$-parameters, and ii) follows from including Eq.~(\ref{epsineq}).
Compared to the data in Eq.~(\ref{rAsym}) we see that the central values of
$\Delta_1$ and $\Delta_2$ currently have opposite signs, disagreeing from
equality by $\sim 2$$\sigma$ when we take into account the theoretical
uncertainty.  The experimental errors are still too large to draw strong
conclusions.

Note that a prediction $|\Delta_1| \approx |\Delta_2|$ was made for the
CP-asymmetries in Ref.~\cite{GR2} based on the expectation that the color
suppressed amplitudes are small.  The CP-sum rule $\Delta_1-\Delta_2+\Delta_3=0$
was discussed in Ref.~\cite{CPsum} (3rd reference) to take into account the
possibly large color suppressed contributions. Given $\zeta^{B\pi}\sim
\zeta_J^{B\pi}$, SCET predicts that the phase of the color suppressed $C_{K\pi}$
amplitude is nearly equal to that of the $T_{K\pi}$ amplitude so the hierarchy
of the asymmetries is actually reinforced by a significant $C_{K\pi}$.  Our
prediction that $|\Delta_1| \gtrsim |\Delta_2|$ with $\Delta_{1,2}$ having equal
signs can also be compared to prediction for the analagous CP-asymmetries in the
QCDF approach~\cite{QCDF} (4th reference). Four different scenarios for the
hadronic parameters were considered S1,S2,S3,S4, and all four sets of model
parameters exhibit the sign correlation. (However all four of the scenarios also
underestimate the size of $|A_{\rm CP}(\pi^+ K^-)|$ by more than a factor of two due
mostly to the fact that the purely perturbative penguin for $K\pi$ is somewhat
small.)

For the branching ratio deviation parameters we have up to smaller terms of
${\cal O}(\epsilon^2)$ that
\begin{align} \label{Rinow}
  R_1 &= \cos(\delta) \big[\epsilon_T^{ew} \!-\! \epsilon \cos(\gamma)  \big] 
  ,\nn\\
 R_2 &= \cos(\delta) \big[ \epsilon_C^{ew} \!-\!\epsilon_T \cos(\gamma)\big] 
  ,\nn\\
 R_3 &= \cos(\delta) \big[ \epsilon_C \cos(\gamma) \!-\! \epsilon^{ew} \big] 
 .
\end{align}
The use of conservative errors on the $\zeta$-parameters leaves too much freedom
to make sign predictions for the $R_i$'s.  However, definite sign predictions
will be possible using Eq.~(\ref{Rinow}) when the $\zeta$ parameters are pinned
down by $B\to\pi$ and $B\to K$ form factor results in the future. Alternatively
accurate measurements of the $R_i$ plus $A_{\rm CP}(K^+\pi^-)$ will determine the
hadronic parameters needed to predict the magnitude of the remaining
$\Delta_i$'s.

In the next section we turn to more direct comparisons of the SCET predictions
with the data by fixing the parameters with the well measured observables and
then predicting the rest.

\subsubsection{$B^- \to \pi^- \bar K^0$ and $B \to \pi^- \bar K^+$}

The amplitude parameters $\hat T_{K \pi}$, $\hat A_{K \pi}$ and $\hat{EW}^C_{K
  \pi}$ are determined in terms of the parameters $\zeta^{B \pi}$ and
$\zeta_J^{B \pi}$ obtained previously from the decays $B \to \pi\pi$. Thus, only
two new parameters are required for the decays $B^- \to \pi^- \bar K^0$ and
$\bar B^0 \to \pi^+K^-$: the magnitude and phase of $P^{K \pi}$.  Since the
ratio of $\lambda_u \hat A_{K \pi} \ll \lambda_c \hat P_{K \pi} \sim 0.001$, one
predicts a negligible CP asymmetry in $\bar B^-\to \pi^- \bar K^0$ in agreement
with the data. The best sensitivity on the two parameters is from ${\rm Br}(B^-
\to \pi^- \bar K^0)$ and $A_{\rm CP}(B^0 \to \pi^+K^-)$. Using these two
observables we find two solutions for $A_{cc}^{K \pi}$ for $\gamma = 83^\circ$
\begin{align}
 10^3\, \hat P^{K\pi} = \left\{ 
 \begin{array}{l} (5.5 \pm 0.1\pm0.1)e^{i(144\pm 8\pm11)^\circ}
  \\ 
  (5.5 \pm 0.1\pm0.1)e^{i(32\pm 7\pm10)^\circ} \end{array} \right.
\end{align}
while for $\gamma = 59^\circ$ we find 
\begin{eqnarray}
 10^3\, \hat P^{K\pi} = \left\{ \begin{array}{l} (5.5 \pm 0.1\pm0.1)e^{i(144\pm 9\pm11)^\circ}\\ (5.5 \pm 0.1\pm0.1)e^{i(36\pm 8\pm10)^\circ} \end{array} \right.
\end{eqnarray}

\begin{figure}[t!]
\begin{center}
\includegraphics[width=9.0cm]{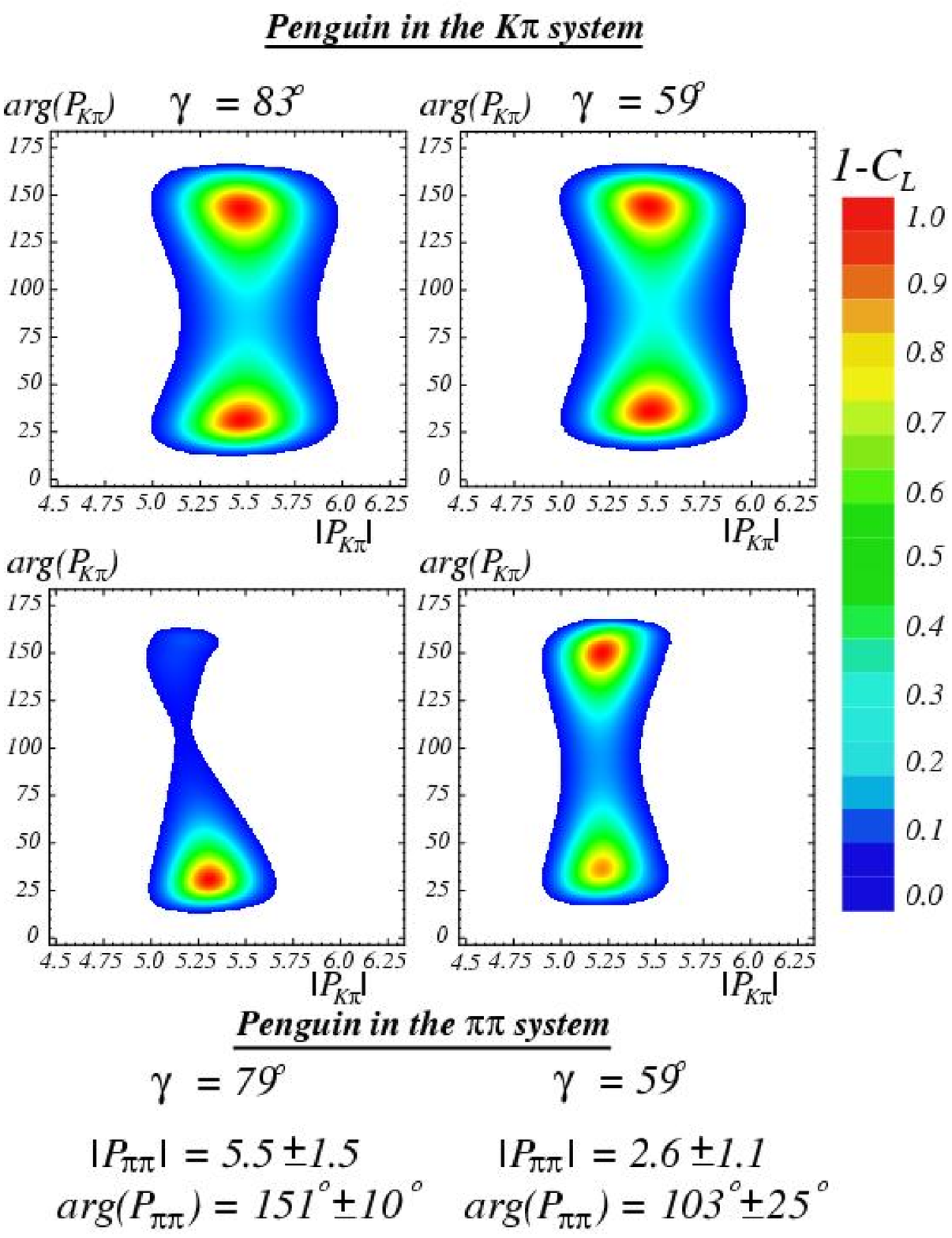}
\caption{Confidence level plots for the complex parameter $A_{cc}^{K \pi}$ for $\gamma=83^\circ$ (left hand side) and $\gamma=59^\circ$ (right hand side). On the top we show the confidence levels without using $Br(\bar B^0 \to \pi^+K^-)$, while the bottom plot includes this branching ratio. We also show the value of $P_{\pi\pi}$, which is idential to the $P_{K \pi}$ in the SU(3) limit. }
\label{fig1}
\end{center}
\end{figure}

The confidence level plot for the magnitude and phase of $P^{K\pi}$ is shown on
the left of Fig.~\ref{fig1}.  For the $\gamma=59^\circ$ result the magnitude
indicates a large SU(3) violating correction at leading order in $\Lambda/E_\pi$
or a large $\Lambda/E_\pi$ correction in the SU(3) limit (which disfavors this
solution). Taking the $\gamma=83^\circ$ we see that of the two solutions the
first has a phase which agrees well with the SU(3) relation to the phase in
$\pi\pi$, while the second phase is quite different.

For $\gamma=83^\circ$ the first solution, however, does not give good agreement
with the third piece of data, the branching ratio ${\rm Br}(B^0 \to \pi^+K^-) =
(18.2 \pm 0.8)\times 10^{-6}$, while the second agrees considerably better. We
find
\begin{eqnarray}
{\rm Br}(\pi^+K^-) = \left\{ 
  \begin{array}{l} (24.0 \pm 0.2\pm1.2)\times 10^{-6}
\\ 
  (21.3 \pm 0.2\pm1.3)\times 10^{-6}  \end{array} \right.
\end{eqnarray}
For $\gamma = 59^\circ$ this branching ratio has much less discriminating power between these two solutions and we find
\begin{eqnarray}
{\rm Br}(\pi^+K^-) = \left\{ \begin{array}{l} 
  (22.5 \pm 0.2\pm1.2)\times 10^{-6}\\
   (22.7 \pm 0.3\pm1.2)\times 10^{-6}  
 \end{array} \right. 
\end{eqnarray}
This is can also be clearly seen in the confidence level plot for $P^{K\pi}$ on
the right of Fig.~\ref{fig1}, where we have included the branching ratio
measurement in the fit. Note, however that both solutions have trouble
explaining the small branching ratio ${\rm Br}(B^0 \to \pi^+K^-)$, making the
large difference in the branching ratios of $B \to \pi^+ K^-$ and $B \to \pi^-
\bar K^0$ quite difficult to explain at LO in the $\Lambda/m_b\ll 1$ limit of QCD.

\subsubsection{Predictions for other $K\pi$ and $K\bar K$ observables}

Using the hadronic parameters extracted from the $B \to \pi \pi$ decays ($\zeta^{B\pi}$, $\zeta_J^{B\pi}$ and $P_{\pi\pi}$), the
value for $P_{K \pi}$ determined from the decays $B^- \to \pi^- \bar K^0$ and
$\bar B^0 \to \pi^- K^+$ decays and independently varying
$\zeta^{BK}+\zeta^{BK}_J = 0.2\pm 0.1$ and $\beta_\pi \zeta^{BK}_J = 0.10\pm
0.05$, we can calculate all the remaining currently measured $K\pi$ observables.
The results are given in Table~\ref{compare} for $\gamma=83^\circ$ and
$\gamma=59^\circ$, respectively. We also show these results in Figs.~\ref{fig2}
and~\ref{fig3}. The data used in the fit are shown in red below the dashed
dividing line while those above the line are predictions.  Note that there is
one more piece of data below the line than there are parameters.

In Fig.~\ref{fig2} we see that $\gamma=83^\circ$ gives a good match to the $B\to
\pi\pi$ data except for the asymmetry $C(\pi^0\pi^0)$. When taking into account
the theoretical error the most striking disagreements are the ${\rm
  Br}(K^-\pi^+)$ at $2.3\sigma$ and the CP-asymmetry $A_{\rm CP}(K^-\pi^0)$ at
$2.6\sigma$. All other predictions agree within the uncertainties. Note that one could demand that $A_{\rm CP}(K^-\pi^0)$ be reproduced, which would imply a negative value of $\zeta_J^{BK}$ (a naive fit for $\gamma = 83^\circ$ gives $\zeta_J^{BK}\sim -0.15$). Note however, that this would imply that both perturbation theory at the intermediate scale $\mu = \sqrt{E \Lambda}$ and SU(3) are badly broken. 

The situation in Fig.~\ref{fig3} with $\gamma=59^\circ$ is similar except that
the theoretical prediction for ${\rm Br}(\pi^+\pi^0)$ moves somewhat. The ${\rm
  Br}(K^-\pi^+)$ deviation is reduced to $1.6\sigma$ and asymmetry $A(K^-\pi^0)$
is still $2.6\sigma$. All other predictions agree within the uncertainties.

\begin{table}[t!]
\begin{tabular}{l|c|c|c}
& Expt. & Theory & Theory \\[-2pt]
& & ($\gamma=83^\circ$) & ($\gamma=59^\circ$) \\[2pt]
\hline
 Data in Fit \\
\hline
$S({\pi^+\pi^-})$ &\ $-0.50\pm 0.12$ \
  & \ $-0.50\pm 0.10 $ & $-0.51 \pm 0.10 $ \ \\
$C({\pi^+\pi^-})$ & $-0.37\pm 0.10$ 
  & $-0.37\pm 0.07$ & $-0.38\pm 0.07 $\\
${\rm Br}(\pi^+ \pi^-)$ & $5.0\pm 0.4$ 
  & $5.0 \pm 2.0$ & $4.6 \pm 1.8 $\\
${\rm Br}(\pi^+\pi^0)$&$5.5\pm 0.6$ 
  & $5.5\pm 2.2$ & $7.3\pm 2.9$ \\
${\rm Br}(\pi^0\pi^0)$&$1.45\pm 0.29 $
  & $1.45\pm 0.58$ & $1.32 \pm 0.53$ \\
${\rm Br}(\bar{K}^0 \pi^-) $ & $24.1\pm 1.3$
  & $24.1\pm 1.2$ &  $24.1\pm 1.2$ \\
$A({ {K}^-\!\pi^+})$ & $-0.115\pm 0.018$
 & $-0.115\pm 0.023$ & $-0.115\pm 0.023$ \\
${\rm Br}(\bar K^0 K^-)$& $1.2\pm0.3$ 
 & $1.2 \pm 0.5 $ &   $1.2 \pm 0.5 $ \\
\hline 
 Predictions \\
\hline
$A({\pi^+ \pi^0})$& $0.01\pm 0.06$ & $\lesssim 0.05$ & $\lesssim 0.05$ \\
$A({\pi^0\pi^0})$&$0.28\pm 0.40$ 
 & $-0.48\pm 0.19 $ & $-0.52\pm 0.27 $ \\
$S(\pi^0\pi^0)$ & & $0.84 \pm 0.23$ & $-0.14\pm 0.22$ \\
${\rm Br}(\pi^0\bar{K}^0)$ & $11.5\pm 1.0$
 & $10.4 \pm 1.1$ & $10.9 \pm 1.2$ \\
${\rm Br}(\pi^+{K}^-)$ & $ 18.9\pm 0.7$
 & $24.0\pm 2.1$ & $22.5\pm 2.1$ \\
${\rm Br}(\pi^0{K}^-)$ &$12.1\pm 0.8$ 
 & $14.3\pm 1.5$ & $12.7\pm 1.4$ \\
$S({\pi^0 K_S})$ & $0.31\pm 0.26$ 
 & $0.77\pm 0.16$ & $0.76\pm 0.16$ \\
$A({\pi^0\! {K}^-})$ & $0.04\pm 0.04$
 & $-0.183\pm 0.075$ & $-0.184\pm 0.076$  \\
$A({\bar{K}^0\!\pi^0})$ & $-0.02\pm 0.13$ 
 & $0.103\pm 0.058$ &  $0.083\pm 0.047$ \\
$A({\pi^-\!\bar{K}^0})$ & $-0.02\pm 0.04 $
 & $< 0.1$ &  $< 0.1 $ \\
${\rm Br}(K^0\bar K^0)$ & $0.96 \pm 0.25 $
 & $1.1 \pm 0.3$ & $1.1\pm 0.3$ \\
${\rm Br}(K^+\! K^-)$ & $0.06\pm 0.12$ 
 & $ \lesssim 0.1$ & $\lesssim 0.1$  \\
$A(\bar K^0 K^-)$ & & $\lesssim 0.2 $ & $\lesssim 0.2 $ \\
$A(\bar K^0 K^0)$ & & $\lesssim 0.2 $ & $\lesssim 0.2 $ \\
\hline
\end{tabular}
{\caption {Comparison of LO predictions versus data as in Figs.3,4. ${\rm Br}$'s are
    in units of $10^{-6}$. The theory
    errors displayed for quantities used in the fit show the relative weight 
    for these observables from power corrections,
    while those for predictions include parameter uncertainty from the fit as
    well as from power corrections. 
  CP asymmetries that are not
  shown in the table are not determined at this order.}\label{compare}}
\end{table}

For $B\to K\bar K$ the amplitude parameters in SCET satisfy $A_{KK} = B_{KK}$
and $E_{KK} = PA_{KK} = EW_{KK} = 0$, and we obtain the prediction
\begin{eqnarray}
{\rm Br}(B^- \to K^- \bar K^0) &=& {\rm Br}(\bar B^0 \to K^0 \bar K^0)
\end{eqnarray}
which agrees well with the latest data, and the expectation that $A_{\rm CP}(B^-
\to K^- \bar K^0)$, $A_{\rm CP}(\bar B^0 \to K^0 \bar K^0)$, and ${\rm Br}(\bar
B^0 \to K^- \bar K^+)$ will be suppressed.

Unfortunately, without the use of SU(3) we do not have enough experimental
information to determine the hadronic parameters required to predict the $B\to
K^0\bar K^0$ absolute branching ratio. It is however interesting to extract the
penguin amplitude and compare with the other channels. We find
\begin{align}
 10^3 |\hat P_{K\bar K}| = 5.3 \pm 0.8
\end{align}
Comparing with the penguin amplitudes extracted in $\pi\pi$ and in $K\pi$ we see
that the combined SU(3) and SCET prediction, $P_{\pi\pi}\sim P_{K\pi}\sim
P_{K\bar K}$, works quite well if $\gamma=83^\circ$.

\section{Conclusions}
\label{sec:conclusions}
\begin{figure}[t!]
\begin{center}
\includegraphics[width=8.5cm]{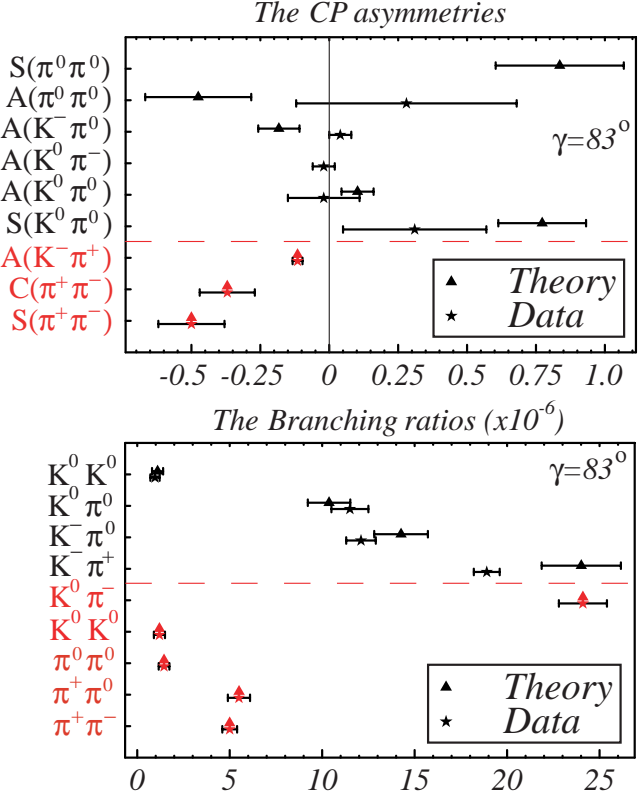}
\caption{Comparison of theory and experiment for all available data in $B \to
  \pi \pi$ and $B \to K \pi$ decays, with $\gamma=83^\circ$. The 8 pieces of data in red
  (below the dashed line) have been used to determine the SCET hadronic parameters 
  $\zeta^{B\pi}$, $\zeta_J^{B\pi}$, $P_{\pi\pi}$, $P_{K\pi}$ and $|P_KK|$, with
  $\zeta^{BK}$ and $\zeta_J^{BK}$ fixed as described in the text. The data above the line are predictions. The CP
  asymmetry in $B^- \to K^0 \pi^-$ is expected to be small, but its numerical 
  value is not predicted reliably.
\label{fig2}}
\end{center}
\end{figure}

\begin{figure}[t!]
\begin{center}
\includegraphics[width=8.5cm]{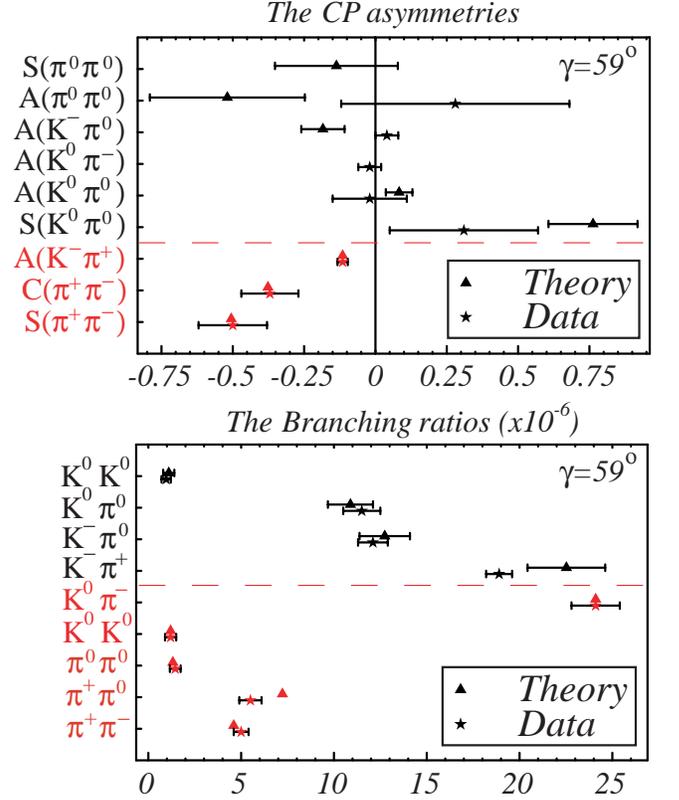}
\caption{Same as Fig.3, but with $\gamma=59^\circ$.
\label{fig3}}
\end{center}
\end{figure}

Decays of $B$ mesons to two pseudoscalar mesons provide a rich environment to
test our understanding of the standard model and to look for physics beyond the
standard model. The underlying electroweak physics mediating these decays are
contained in the Wilson coefficients of the electroweak Hamiltonian as well as
CKM matrix elements. In order to test cleanly the standard model predictions for
these short distance parameters, one requires a good understanding of the
QCD matrix elements of the effective operators, which can not be calculated
perturbatively.

At the present time, there are 5 well measured (with $< 100\%$ uncertainty)
observables in $B \to \pi \pi$, 5 in $B \to K \pi$ and 2 in $B \to K K$. Using
only isospin symmetry (with corrections suppressed by $m_{u,d}/\Lambda$), 
the number of hadronic parameters required to describe these decays
is 7, 11 and 11, respectively. The number of hadronic parameters can be reduced
by two in the $\pi \pi$ system, if one drops the two operators $O_7$ and $O_8$,
which have small Wilson coefficients in the standard model. If one is willing to
take SU(3) (an expansion in $m_s/\Lambda$) as a good symmetry of QCD, the combined $B \to \pi\pi/K\pi$ system is
described by 15 parameters, while the $B \to KK$ system adds another 4
parameters. Neglecting $O_7$ and $O_8$ with SU(3) reduces the number of
parameters in the $\pi\pi/K\pi/K\bar K$ system to 15. Thus, at the present time
there are more hadronic parameters than there are well measured observables.

In this paper we have studied these decays in a model independent way using
SCET. This analysis exploits that the hadronic scale $\Lambda$ in QCD is much
smaller than both in the large mass of the heavy quark and the large energy of
the two light mesons.  It follows that at leading order in the power expansion
in $\lqcd/Q$, where $Q \sim m_b,E$, and using SU(2), there are four hadronic
parameters describing $B \to \pi \pi$, five additional parameters describing $B
\to K \pi$ and three additional parameters describing $B \to KK$. In the limit
of exact SU(3) the four parameters describing $B \to \pi \pi$ are enough to
describe all of these $B \to PP$ decays in SCET.

In SCET the electroweak penguin operators $O_{7,8}$ can be included without
adding additional hadronic parameters.  One can use the 5 pieces of well
measured $\pi\pi$ data to determine the four hadronic parameters and the weak angle
$\gamma$~\cite{gammafit}, and with the current data one finds $\gamma=83^\circ
\pm 8^\circ \pm 2^\circ$.  This is still consistent with the direct measurement
of this angle from $B\to DK$~\cite{HFAG}, but is currently in conflict with the
value of $\gamma$ from a global fit of the unitarity triangle at the $2\sigma$
level.  It is too early to tell if this implies larger than expected power
corrections in SCET or might be a first hint at new physics.  When we proceed to
analyze the decays $B \to K \pi$, we thus perform our analysis both for $\gamma
= 83^\circ$ and $\gamma = 59^\circ$. For both of these values the direct CP
asymmetry in $B \to \pi^0 \pi^0$ is predicted to have the opposite sign from the
measured value, but is still consistent at the $2\sigma$ level.

Moving on to $B \to K \pi$ decays, we analyzed the uncertainty in the Lipkin sum
rule~\cite{Lipkin} for branching fractions and the CP-sum rule~\cite{CPsum}
for rescaled CP-asymmetries as defined in Eqs.~(\ref{R},\ref{rAsym}), giving our
result in Eqs.~(\ref{sumerrors1},\ref{sumerrors2}). The CP-sum rule was found to
be particularly accurate due to a suppression by three small parameters in SCET.
The Lipkin sum rule is second order in small parameters and has a theoretical
precision that also makes it an interesting observable. We conclude that both
the Lipkin and CP-sum rules will provide very robust methods for testing the
$K\pi$ data as the experimental errors decrease in the future.

Using the expectation that the hadronic parameters $\zeta^{BM}$ and
$\zeta_J^{BM}$ in the factorization theorem are positive, we showed that the rescaled asymmetry
$\Delta_1(\pi^0K^-)$ should have the same sign and larger magnitude than the
rescaled asymmetry $\Delta_2(\pi^-K^+)$ which is well measured. This prediction
is in conflict with the current data by $\sim 2\sigma$. Other sign and magnitude
predictions are discussed in section~\ref{sec:CPsign}.

The SCET amplitude formulas predict that in addition to the $\pi\pi$ parameters
already determined, only the complex $K\pi$ penguin amplitude is required to
describe the decays $B^- \to \pi^- \bar K^0$ and $B \to \pi^+ K^-$. This happens
because they involve $\zeta^{B\pi}$ and $\zeta_J^{B\pi}$, but do not involve
$\zeta^{B\bar K}$ or $\zeta^{B\bar K}_J$.  The well know prediction of a small
CP asymmetry for $B^- \to \pi^- \bar K^0$ is reproduced in SCET. The large
difference in ${\rm Br}(B \to \pi^+ K^-)$ and ${\rm Br}(B^- \to \pi^- \bar K^0)$
is difficult to explain in the standard model with SCET. The $\gamma = 59^\circ$
solution is not preferred by the combined SU(3)+SCET limit which predicts
$P_{K\pi}\simeq P_{\pi\pi}$. These amplitudes agree well for $\gamma=83^\circ$.

Given the current uncertainties in the data, the remaining two hadronic
parameters $\zeta^{BK}$ and $\zeta_J^{BK}$ can not yet be determined reliably. This also means that predictions for
the remaining rates do not depend too sensitively on these parameters. Fixing their
values to be close to those preferred by SU(3), but with 50\% 
uncertainty,  we obtained predictions for the
remaining observables in Figures~\ref{fig2} and~\ref{fig3}. 

Finally, the decays $B \to KK$ require two additional hadronic parameters, which
can only be determined once better data for both rates and CP asymmetries become
available for these decays. One prediction of SCET, namely that ${\rm Br}(B \to
K^0 \bar K^0) = {\rm Br}(B^- \to K^- \bar K^0)$ is well satisfied by the current
data. In the SU(3) limit one expects that $ P_{K\pi}\sim P_{K\bar K}$, and this
result is in good agreement with the data.

In conclusion, several predictions of SCET work rather well, while for others
there are discrepancies with the current data. It is too early to tell if the
disagreements between theory and data are due to statistical fluctuations, to
larger than expected power corrections or if they reveal a first glimpse of
physics beyond the standard model. To answer this question, the experimental
uncertainties need to be reduced and the convergence of the SCET expansion of
QCD for nonleptonic decays has to be tested further both with nonleptonic and with semileptonic data~\cite{agrs,Hillf}.

\begin{acknowledgments}
  This work was supported by the Director, Office of Science, Office of High
  Energy, Division of High Energy Physics under Contract DE-AC03-76SF00098
  (C.B.), DOE-ER-40682-143 and DEAC02-6CH03000 (I.R.), the Office of Nuclear
  Science and cooperative research agreement DF-FC02-94ER40818 (I.S.), and the
  DOE OJI program and Sloan Foundation (I.S.).  We would like to thank the
  Institute for Nuclear theory (INT), where part of this work was performed, and
  D.~Pirjol and Z.~Ligeti for helpful suggestions.
\end{acknowledgments}

\begin{appendix}

\section{Operators and Matrix Elements in SCET}\label{appSCET}

At the scale $\mu\simeq m_b$ the Hamiltonian in Eq.~(\ref{Hw}) is matched onto
operators in SCET. For the first two orders in the power expansion
\begin{align} \label{match}
 H_W \!\!&=  &\!\! \frac{2G_F}{\sqrt{2}} \sum _{n,\bn} \bigg\{ 
  \sum_i \int [d\omega_{j}]_{j=1}^{3}
       c_i^{(f)}(\omega_j)  Q_{if}^{(0)}(\omega_j) \nn\\ 
 && \hspace{-1cm}
  + \sum_i \int [d\omega_{j}]_{j=1}^{4}  b^{(f)}_i(\omega_j) 
  Q_{if}^{(1)}(\omega_j) 
  + {\cal Q}_{c\bar c} + \ldots \bigg\} \,,
\end{align}
The Wilson coefficients $c_i$ and $b_i$ are the Wilson coefficients that appear
in Eqs.~(\ref{pipiamps}--\ref{KKamps}).  The operators for the $\Delta S=0$
transitions are~\cite{pipiChay,bprs}
\begin{eqnarray} \label{Q0}
  Q_{1d}^{(0)} &=&  \big[ \bar u_{n,\omega_1} \bnslash P_L b_v\big]
  \big[ \bar d_{\bn,\omega_2}  \nslash P_L u_{\bn,\omega_3} \big]
  \,,  \\
  Q_{2d,3d}^{(0)} &=&  \big[ \bar d_{n,\omega_1} \bnslash P_L b_v \big]
  \big[ \bar u_{\bn,\omega_2} \nslash P_{L,R} u_{\bn,\omega_3} \big]
   \,,\nn \\
  Q_{4d}^{(0)} &=&  \big[ \bar q_{n,\omega_1} \bnslash P_L b_v \big]
  \big[ \bar d_{\bn,\omega_2} \nslash P_{L}\, q_{\bn,\omega_3} \big]
   \,, \nn \\
  Q_{5d,6d}^{(0)} &=&  
  \big[ \bar d_{n,\omega_1} \bnslash P_L b_v \big]
  \big[ \bar q_{\bn,\omega_2} \nslash P_{L,R} q_{\bn,\omega_3} \big]
  \,, \nn
\end{eqnarray}
and
\begin{eqnarray}
  Q_{1d}^{(1)} \!\!&=&\!\! \frac{-2}{m_b} 
     \big[ \bar u_{n,\omega_1}\, ig\,\slash\!\!\!\!{\cal B}^\perp_{n,\omega_4} 
     P_L b_v\big]
     \big[ \bar d_{\bn,\omega_2}  \nslash P_L u_{\bn,\omega_3} \big] 
     \,, \nn \\
  Q_{2d,3d}^{(1)} &=&  \frac{-2}{m_b}  
     \big[ \bar d_{n,\omega_1} \, ig\,\slash\!\!\!\!{\cal B}^\perp_{n,\omega_4} 
     P_L b_v \big]
     \big[ \bar u_{\bn,\omega_2} \nslash P_{L,R} u_{\bn,\omega_3} \big]
      \,,\nn \\
  Q_{4d}^{(1)} &=&  \frac{-2}{m_b} 
     \big[ \bar q_{n,\omega_1} \, ig\,\slash\!\!\!\!{\cal B}^\perp_{n,\omega_4} 
     P_L b_v \big]
     \big[ \bar d_{\bn,\omega_2} \nslash P_{L}\, q_{\bn,\omega_3} \big]
      \,,\nn \\
  Q_{5d,6d}^{(1)} &=& \frac{-2}{m_b} 
    \big[ \bar d_{n,\omega_1} \, ig\,\slash\!\!\!\!{\cal B}^\perp_{n,\omega_4} 
     P_L b_v \big]
    \big[ \bar q_{\bn,\omega_2} \nslash P_{L,R} q_{\bn,\omega_3} \big]
      \,, \nn\\
  Q_{7d}^{(1)} \!\!&=&\!\! \frac{-2}{m_b} 
   \big[ \bar u_{n,\omega_1}\, ig\,{\cal B}^{\perp\, \mu}_{n,\omega_4} 
    P_L b_v\big]
   \big[ \bar d_{\bn,\omega_2}  \nslash \gamma^\perp_\mu P_R u_{\bn,\omega_3} \big] 
      \,,\nn\\
  Q_{8d}^{(1)} \!\!&=&\!\! \frac{-2}{m_b} 
   \big[ \bar q_{n,\omega_1}\, ig\,{\cal B}^{\perp\, \mu}_{n,\omega_4} 
    P_L b_v\big]
   \big[ \bar d_{\bn,\omega_2}  \nslash \gamma^\perp_\mu P_R q_{\bn,\omega_3} \big] 
      \,. \nn
\end{eqnarray}
The $\Delta S=1$ operators $Q_{is}^{(0)}$ are obtained by swapping $\bar d\to
\bar s$.  The ``quark'' fields with subscripts $n$ and $\bn$ are products of
collinear quark fields and Wilson lines with large momenta $\omega_i$. We have
defined
\begin{align}
  \bar u_{n,\omega} &= [ \bar\xi_n^{(u)} W_n\, \delta(\omega\!-\!
\bn\mcdot\cP^\dagger) ]\,, \\
 ig\,{\cal B}^{\perp\,\mu}_{n,\omega} &= \frac{1}{(-\omega)}\, 
 \big[ W^\dagger_n [ i\bn\mcdot D_{c,n} , i D^\mu_{n,\perp} ] W_n 
  \delta(\omega-\bnP^\dagger) \big] \nn 
\end{align}
where $\bar\xi_n^{(u)}$ creates a collinear up-quark moving along the $n$
direction, or annihilates an antiquark.  The $b_v$ field is the standard HQET
field.  For a complete basis we also need operators with octet bilinears,
$T^A\otimes T^A$, but their matrix elements vanish at LO.  The operators
$Q_{7d}^{(1)}$ and $Q_{8d}^{(1)}$ also do not contribute at LO~\cite{bprs}, see also~\cite{Kagan}.

The leading order factorization theorem in Eq.~(\ref{A0newfact}) is generated by
time ordered products of both the operators $Q^{(0)}$ and $Q^{(1)}$ with
insertions of a subleading Lagrangian.  T-products with $Q^{(0)}$ contribute to
terms with $\zeta^{BM}$ and T-products with $Q^{(1)}$ contribute to those with
$\zeta_J^{BM}$.   It is convenient to define 
\begin{align}
  \tilde Q_i^{(0)} &= \big[ \bar q^{i}_{n,\omega_1} \bnslash P_L b_v \big] \,,
  \\
  \tilde Q_i^{(1)} &= \frac{-2}{m_b} \big[ \bar q^i_{n,\omega_1}\,
  ig\,\slash\!\!\!\!{\cal B}^\perp_{n,\omega_4} P_L b_v\big] \,,\nn\\
  \tilde Q_i^\bn &= \bar q^i_{\bn,\omega_2} \nslash P_{L,R} q^{\prime
    i}_{\bn,\omega_3} \,.
\end{align}
In $\tilde Q_i^{(0,1)}$ the flavor of the $\bar q^i_{n,\omega_1}$ terms matches
that of the first bilinear in Eq.~(\ref{Q0}). In $\tilde Q_i^\bn$ the flavor of
$\bar q^i$ and $q^{\prime i}$ match those in the second bilinear of
Eq.~(\ref{Q0}), and we have $P_R$ for $i=3,6$ and $P_L$ otherwise.  The
contributions to $B\to M_1 M_2$ at LO are all from $\tilde Q_i^\bn$ times the
time-ordered products
\begin{align} \label{Tproducts}
  T_1^i &=\! \mbox{\large $\int$} d^4y\,
    T \big[\tilde Q^{(0)}_i(0)\: i{\cal L}^{(1)}_{\xi_n q}(y)\big] \nn\\
    & \ \  + \mbox{\large $\int$} d^4y\, d^4y'\,T 
   \big[\tilde Q^{(0)}_i(0) \: i{\cal L}^{(1)}_{\xi_n q}(y) \: i{\cal
    L}_{\xi_n\xi_n}^{(1)}\!(y') \big] \!\nn\\
  & \ \ + \mbox{\large $\int$} d^4y\, d^4y'\,T 
   \big[\tilde Q^{(0)}_i(0) \: i{\cal L}^{(1)}_{\xi_n q}(y) \: 
   i{\cal L}_{cg}^{(1)}(y')\} \big]\nn \\[4pt]
  & \ \ 
     +  \mbox{\large $\int$} d^4y\, 
    T \big[\tilde Q_i^{(0)}(0),i{\cal L}^{(1,2)}_{\xi_n q}(y) \big], \ \nn\\
  T_2^i(z) &=\! \mbox{\large $\int$} d^4y \:
    T \big[\tilde Q_i^{(1)}(0),i{\cal L}^{(1)}_{\xi_n q}(y) \big] ,
\end{align} 
where $z$ and $1-z$ are the momentum fractions carried by the collinear quark
and gluon field in $\tilde Q_i^{(1)}$.  Here $T_1$ and $T_2$ are exactly the
same T-products that occur in the heavy-to-light form factors~\cite{bps5}.  In
addition we have operators/T-products whose matrix elements give $A_{cc}$ (see
the appendix of Ref.~\cite{diff1} for further discussion of these
contributions).  Using the collinear gluon fields defined in Ref.~\cite{bps6}
the Lagrangians in Eq.~(\ref{Tproducts}) are
\begin{eqnarray} \label{Lxxnew}
{\cal L}_{\xi\xi}^{(1)} 
  &=&  \big(\bar \xi_n  W\big)\,  i\Dslash^\perp_{us} 
  \frac{1}{\bnP} \big(W^\dagger 
  i \Dslash^\perp_c  \frac{\bnslash}{2} \xi_n \big)
  +\mbox{h.c.}\,, \\
  {\cal L}^{(1)}_{\xi q} &=&   \bar\xi_n \: \frac{1}{i\bn\mcdot D_c}\: 
 ig\, \Bslash_\perp^c W  q_{us} \mbox{ + h.c.}\,,\nn\\
    {\cal L}^{(2)}_{\xi q} &=&  \bar\xi_n \frac{\bnslash}{2}
     \frac{1}{i\bn\mcdot D_c}\: 
     ig\,  n\mcdot M \, W \, q_{us}  \nn\\
   && +  \bar\xi_n \frac{\bnslash}{2} 
  i\Dslash_\perp^{\,c} \frac{1}{(i\bn\mcdot D_c)^2}\:   ig\, \Bslash_\perp^c W 
  \: q_{us}   \mbox{ + h.c.} \,, \nn \\
 {\cal L}_{cg}^{(1)} &=& \frac{2}{g^2}\: {\rm tr} 
  \Big\{ \big[i {D}_0^\mu , iD_c^{\perp\nu} \big] 
         \big[i {D}_{0\mu} , W iD_{us\,\nu}^\perp W^\dagger \big] \Big\}
  \,, \nn
\end{eqnarray}
where $i D_0^\mu = i{\cal D}^\mu + g A_{n}^\mu$.

In this paper we only used this factorization at the scale $m_b$, so the
hadronic paramaters are defined by matrix elements of $T_1$ and $T_2$ and the
$\bn$-collinear operator, namely
\begin{align}
  & \langle M_n | T_1^i | B\rangle = {\cal C}_i(B,M) \: m_B\: \zeta^{BM} \,,\\
  & \langle M_n | T_2^i(z) | B\rangle = {\cal C}_i(B,M) \: m_B\: \zeta_J^{BM}(z)
  \,,
  \nn\\
  & \langle M_\bn | \tilde Q_i^\bn |
  0\rangle = {\cal C}_i^\prime(B,M) \: m_B\: f_M\: \phi_{M}(u)\nn
\end{align}
where $u$ and $1-u$ are momentum fractions for the quark and antiquark
$\bn$-collinear fields.  Here ${\cal C}^i(B,M)$ and ${\cal
  C}_i^\prime(B,M)$ are simple Clebsch-Gordan coefficients. Putting the
pieces together we have
\begin{align}
 A &=  \langle M_1 M_2 | H_W | \bar B \rangle \\
 &= \frac{2 G_F m_B^2}{\sqrt{2}} \sum_i {\cal C}_i(B,M_1) \:
  {\cal C}_i^\prime(B,M_2)\: f_{M_2}  \nn \\
 &\ \ \times \bigg[ 
  \int_0^1\!\! du dz \: b_i(u,z) \zeta_J^{B M_1}(z) \phi_{M_2}(u)
  \nn\\
 &\qquad +  \zeta^{B M_1} \int_0^1\!\! du\: c_i(u) \phi_{M_2}(u)  
  \bigg] + (1\leftrightarrow 2) \,
  \nn\\
 & + A_{c\bar c}^{M_1 M_2} \,.\nn
\end{align}
This result was used to obtain Eq.~(\ref{pipiamps}--\ref{KKamps}) where the
relevant combinations of ${\cal C}^i\: {\cal C}^{\prime i}$ coefficients can be
read off from Table~I of Ref.~\cite{bprs} (and do not asssume isospin symmetry).
Here $A_{c\bar c}^{M_1 M_2}$ contains Clebsch Gordan coefficients if for example
SU(2) is used to relate these parameters in different channels. For amplitudes
with no penguin contribution we have $A_{c\bar c}^{M_1 M_2}=0$.

\section{Relationship between our amplitude parameterization and graphical amplitudes}

In this appendix we show the relationship between the amplitude parameters defined in Eqs.~(\ref{pipigeneral})-(\ref{KKgeneral}) and the graphical amplitudes defined in~\cite{graphical1,graphical2}. These relations are useful, since one can immediately read off SU(3) relations between different amplitudes, since the graphical amplitudes are SU(3) invariant. 
Note that while the amplitude paramters on the right hand side of equations (\ref{B1},\ref{B2}, \ref{B3})
have the same name for the different processes, $\pi\pi$, $K\pi$, and $K\bar K$, they are only equal in the SU(3) limit. 

The relations for the amplitude parameters in  $B \to \pi \pi$ are
\begin{eqnarray} \label{B1}
\hat T_{\pi\pi} &=& T + P_{ut}+E+PA_{ut}  + EW^C + \frac{EW^A}{2} - \frac{EW^E}{2}\nn\\
&& - \frac{EW^P}{2}-\frac{EW^{PA}}{2} \nn\\
\hat C_{\pi\pi} &=& C-P_{ut} -E-PA_{ut}+ \frac{3 EW^T}{2} + \frac{EW^C}{2} \nn\\ 
&&  +\frac{EW^P}{2}+\frac{EW^{PA}}{2}+\frac{EW^E}{2}-\frac{EW^A}{2} \nn\\
\hat P_{\pi\pi} &=& P_{ct}+PA_{ct} + EW^C + \frac{EW^A}{2} - \frac{EW^E}{2}\nn\\
&&-\frac{EW^P}{2} - \frac{EW^{PA}}{2}\nn\\
\hat{EW}^T_{\pi\pi} &=& \frac{3}{2} (EW^T + EW^C)\,.
\end{eqnarray}
The amplitude parameters  for $B \to K \pi$ decays can be written in terms of graphical amplitudes as follows:
\begin{eqnarray} \label{B2}
\hat T_{K \pi} &=& T+P_{ut}+EW^C-\frac{EW^P}{2}-\frac{EW^E}{2}\nn\\ 
\hat C_{K \pi} &=& C-P_{ut}+\frac{3EW^T}{2}+\frac{EW^C}{2}\nn\\
&&+\frac{EW^P}{2} + \frac{EW^E}{2}\nn\\ 
\hat P_{K \pi} &=& P_{ct} + EW^E - \frac{EW^C}{2} - \frac{EW^P}{2} \nn\\ 
\hat A_{K \pi} &=& P_{ut}+A+ EW^E-\frac{EW^P}{2}-\frac{EW^C}{2}\nn\\ 
\hat{EW}^T_{K \pi} &=& \frac{3}{2} ( EW^T+EW^C) \nn\\
\hat{EW}^C_{K \pi} &=& \frac{3}{2} ( EW^C-EW^E) \,,
\end{eqnarray}
Finally,for $B \to KK$ decays we find
\begin{eqnarray} \label{B3}
\hat A_{KK} &=& P_{ut}+A-\frac{EW^C}{2}+EW^E-\frac{EW^P}{2}\nn\\
\hat B_{KK} &=& P_{ut}+PA_{ut}-\frac{EW^C}{2}-EW^A\nn\\
\hat E_{KK} &=& -E- PA_{ut}-\frac{EW^A}{2}+\frac{EW^{PA}}{2}\nn\\
\hat P_{KK} &=& P_{ct}-\frac{EW^C}{2}+EW^E-\frac{EW^P}{2}\nn\\
\hat {PA}_{KK} &=& PA_{ct} +\frac{EW^A}{2}-\frac{EW^{PA}}{2} \nn\\
\hat {EW}_{KK} &=& -\frac{3EW^A}{2} -\frac{3EW^E}{2} \,.
\end{eqnarray}

\end{appendix}

\end{document}